\documentclass[aps, prd, twocolumn, lengthcheck, superscriptaddress, 
nofootinbib]{revtex4-1}

\usepackage{amsmath}
\usepackage{graphicx}
\usepackage{color}
\usepackage{times}
\usepackage{inputenc}
\usepackage{bm}
\usepackage{multirow}
\usepackage{float}
\usepackage{url}
\usepackage{natbib}
\usepackage{tensor}
\usepackage[colorlinks=true,citecolor=blue,urlcolor=blue,linkcolor=red]{hyperref}
\usepackage[normalem]{ulem}
\usepackage{subfigure}

\def\fm3{\;\text{fm}^{-3}}

\begin{document}
\title{
Quarkyonic matter with chiral symmetry restoration}

\author{Bikai Gao}
\email{gaobikai@hken.phys.nagoya-u.ac.jp}
\affiliation{Department of Physics, Nagoya University, Nagoya 464-8602, Japan}

\author{Masayasu Harada}
\email{harada@hken.phys.nagoya-u.ac.jp}
\affiliation{Kobayashi-Maskawa Institute for the Origin of Particles and the Universe, Nagoya University, Nagoya, 464-8602, Japan}
\affiliation{Department of Physics, Nagoya University, Nagoya 464-8602, Japan}
\affiliation{Advanced Science Research Center, Japan Atomic Energy Agency, Tokai 319-1195, Japan}

\date{\today}

\begin{abstract}
We present a novel unified approach to describe the dense symmetric nuclear matter by combining the quarkyonic matter framework with the parity doublet model. This integration allows for a consistent treatment of the transition from hadronic to quark degrees of freedom while incorporating chiral symmetry restoration effects. Our model introduces a chiral invariant mass for both baryons and constituent quarks, enabling a smooth crossover between hadronic and quark matter in symmetric nuclear matter. We derive the equation of state (EOS) for this hybrid system and investigate its thermodynamic properties. The model predicts a gradual onset of quark degrees of freedom at high densities while maintaining aspects of confinement. 
\end{abstract}

\maketitle 

\section{Introduction}
Recent breakthroughs in multi-messenger astronomy, including the detection of massive neutron stars exceeding two solar masses and the observation of gravitational waves from neutron star mergers\cite{Fonseca:2016tux,LIGOScientific:2017vwq,LIGOScientific:2017ync,Miller:2021qha,Riley:2021pdl,Fonseca:2021wxt,Vinciguerra:2023qxq}, have placed unprecedented constraints on the equation of state (EOS) of ultra-dense matter. These observations have challenged our understanding of nuclear physics at high densities and temperatures, posit stringent constraints on our theoretical models.

Despite these advancements, a consistent description of matter transitioning from nuclear to quark degrees of freedom remains a significant problem. This transition region around $2n_0$ to $5n_0$($n_0$: saturation density), where the fundamental degrees of freedom shift from hadrons to quarks, is particularly challenging to model due to the non-perturbative nature of quantum chromodynamics (QCD) in this regime. Traditional approaches often rely on separate descriptions for hadronic and quark matter, with an abrupt phase transition between the two\cite{Lenzi:2012xz,Benic:2014jia,Zdunik:2012dj,Alford:2013aca,Gartlein:2023vif}. However, recent theoretical evidence suggests that the transition may be a more gentle process, possibly involving a crossover\cite{Masuda:2012kf,Kojo:2014rca,Baym:2017whm,Baym:2019iky,Kojo:2021wax,Gao:2022klm,Minamikawa:2023eky}.

\begin{figure}[htbp]\centering
\includegraphics[width=1\hsize]{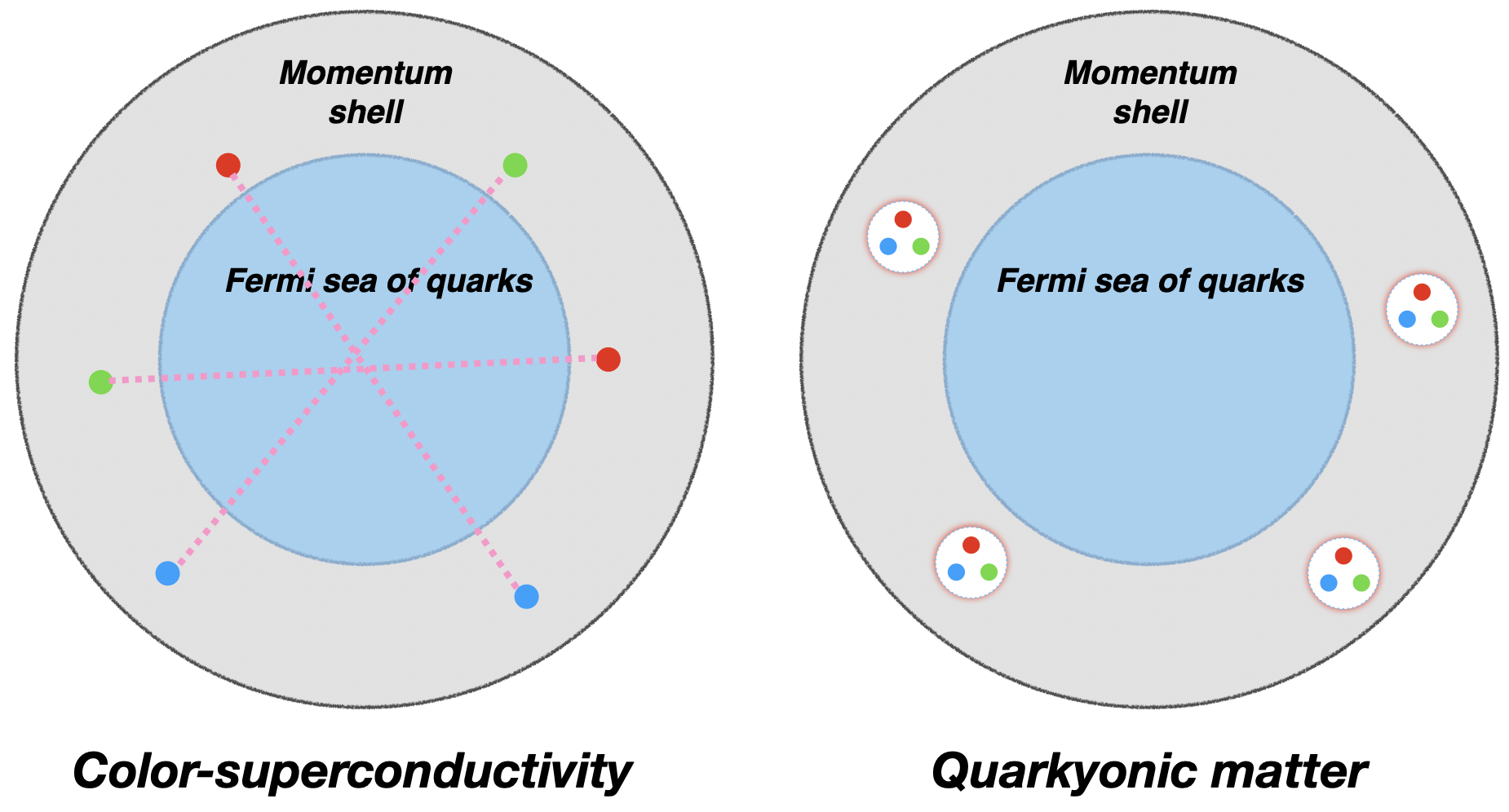}
\caption{Schematic plot showing the structure of Fermi sea for color-superconductivity with two particle correlation (left panel) and the quarkyonic description with three particle correlation (right panel). }
\label{fig_quarkyonic}
\end{figure}

The quarkyonic matter description\cite{McLerran:2007qj,McLerran:2008ua,Hidaka:2008yy,Fukushima:2015bda,Duarte:2021tsx,Kojo:2021ugu,McLerran:2018hbz,Jeong:2019lhv,Sen:2020peq,Fujimoto:2023mzy} offers an intriguing picture of the transition from hadronic to quark matter. The basic concept of quarkyonic matter is that at sufficiently high baryon chemical potential, the degrees of freedom inside the Fermi sea can be treated as quarks, while confining forces remain important only near the Fermi surface. Nucleons emerge through correlations between quarks at the surface of the quark Fermi sea at high densities. This phenomenon is somewhat analogous to Cooper pairing\cite{Bailin:1983bm,Berges:1998rc,Rajagopal:2000wf,Alford:2001dt,Alford:2007xm,Baym:2017whm} in fermionic systems as shown in Fig.~\ref{fig_quarkyonic}. The key distinction is that in quarkyonic matter, quarks are confined into baryons which are colorless. However, the diquarks in Cooper pairing have color. In both theoretical frameworks, they are emphasizing the importance of interactions near the Fermi surface.

The parity doublet model (PDM)\cite{Detar:1988kn,Jido:2001nt} provides a natural framework to describe the chiral properties of baryons. Traditionally, the origin of nucleon mass has been primarily attributed to spontaneous chiral symmetry breaking. For instance, QCD sum rules\cite{Ioffe:1981kw,Dominguez:1986aa} have demonstrated that the nucleon mass is proportional to the quark-antiquark condensate $\langle\bar{q}q \rangle$, suggesting that in the chiral limit, the nucleon mass should approach zero.
However,  the PDM introduces a novel concept: chiral invariant mass $m_0$ which is insensitive to chiral symmetry breaking\cite{Detar:1988kn,Jido:2001nt,Dexheimer:2007tn,Sasaki:2010bp, Motohiro,Nishihara:2015fka,Minamikawa:2023ypn,Gao:2024mew,HATSUDA198911,GALLAS201113,Marczenko:2024jzn,Yasui:2024dbx,Koch:2023oez}. This  feature of the model is particularly intriguing as it offers a mechanism for the origin of baryon masses that is partially independent of the chiral condensate. Lattice QCD simulations have also provided further support for this concept\cite{Aarts:2015mma,Aarts:2017rrl,Aarts:2017iai,Aarts:2018glk}.
In finite density systems, the PDM predicts a gradual restoration of chiral symmetry, with the masses of chiral partners becoming degenerate as the density increases. This behavior could have significant consequences for the EOS of dense matter, potentially affecting the structure and properties of neutron stars\cite{universe5080180,PhysRevC.100.025205,Mukherjee:2017jzi,Minamikawa:2020jfj,Marczenko:2021uaj,Marczenko:2022hyt,Kong:2023nue,Gao:2024chh,Gao:2024lzu,Eser:2024xil}.

In this work, we propose a unified approach that combines the quarkyonic matter framework with the parity doublet model (PDM). This integration aims to provide a more comprehensive description of dense nuclear matter, addressing both confinement properties and chiral symmetry aspects. Our approach is related to the mass origin of constituent quark\cite{DeRujula:1975qlm,Isgur:1978xj,Manohar:1983md,Scavenius:2000qd}, a phenomenological framework that has successfully described many aspects of hadron spectroscopy and static properties. In the constituent quark model, hadrons are composed of constituent quarks, quasi-particles with effective masses much larger than their current quark masses in QCD. For up and down quarks, these effective masses are typically about 300-350 MeV, compared to their current quark masses of only a few MeV. These large effective masses are traditionally attributed to quark-gluon interactions and, crucially, to the phenomenon of dynamical chiral symmetry breaking. We extend this model by incorporating the concept of chiral invariant mass, similar to the PDM. In our framework, the constituent quark mass is divided into two components: a chiral invariant part and a part generated by chiral symmetry breaking. This structure implies that quarks, and consequently hadrons, will retain a non-zero mass even if chiral symmetry is fully restored. Such an approach provides a mechanism for hadrons to maintain some of their properties in environments where chiral symmetry is partially restored, such as in hot or dense nuclear matter.

Our unified approach offers several advantages over previous models. It provides a smooth transition from hadronic to quark degrees of freedom, avoiding the discontinuities often present in traditional hybrid star models. The model also incorporates chiral symmetry restoration effects in a way that is consistent with our expectations from QCD. Furthermore, it offers a more realistic description of baryon and quark masses in dense matter, taking into account both confinement and chiral symmetry aspects.
By combining the PDM and the quarkyonic picture, we aim to provide a more comprehensive description of dense nuclear matter that respects both the confinement properties suggested by the quarkyonic matter picture and the chiral symmetry aspects captured by the parity doublet model. This unified approach may offer new insights into several long-standing puzzles in nuclear physics and astrophysics, such as the nature of the quark-hadron transition and the properties of the densest matter created in heavy-ion collisions. We hope that the future extension of our model has the potential to make predictions for various neutron star observables. These include the mass-radius relationship, tidal deformabilities, and cooling behavior. By comparing these predictions with current and future observations, we can test the validity of our model and potentially uncover evidence for exotic states of matter in neutron star interiors.

This paper is organized as follows: In sec.\ref{sec_formulation}, we explain the formulation of
present model. The main results of the analysis are shown in Sec.~\ref{sec_results} and Sec.~\ref{sec_constituent}. Finally, we show the summary and discussions in  Sec.~\ref{sec_summary}.

\maketitle 
\section{Formulalion}
\label{sec_formulation}
Here, we explain the construction of the relativistic mean-field model based on the parity doublet structure and the quarkyonic description. Previous research in Ref. \cite{McLerran:2018hbz}  studied the impacts of the quarkyonic picture with non-interacting matter with constant nucleon mass and constant constituent quark mass. In this research, we would like to study a more sophisticated case which includes the interaction and also the parity doublet structure. In the parity doublet framework, the excited nucleon $N(1535)$ with negative parity is regarded as the chiral partner of the ground state nucleon $N(939)$ with positive parity. Following Refs.\cite{Gao:2024chh,Minamikawa:2020jfj}, we write the thermodynamic potential in PDM with $N_f=2$ as
\begin{equation}
\begin{aligned}
\Omega_{\mathrm{PDM}}=& V(\sigma)-V_0-\frac{1}{2} m_{\omega}^{2} \omega^{2}\\
&-\frac{1}{2} m_{\rho}^{2} \rho^{2} -\lambda_{\omega \rho}\left(g_{\omega} \omega\right)^{2}\left(g_{\rho} \rho\right)^{2} + \Omega_F\\
\Omega_{F}=&-2 \sum_{i=+,-} \sum_{\alpha=p, n} \int^{k_{f}} \frac{\mathrm{d}^3 \mathbf{p}}{(2 \pi)^{3}}\left(\mu_{\alpha}^{*}-E_{\mathrm{p}}^{i}\right).
\end{aligned}
\label{Omega_PDM}
\end{equation}
with  the potential $V(\sigma)$ is given  by
\begin{align}
V(\sigma) &= -\frac{1}{2}\bar{\mu}^{2}\sigma^{2} + \frac{1}{4}\lambda_4 \sigma^4 -\frac{1}{6}\lambda_6\sigma^6 - m_{\pi}^{2} f_{\pi}\sigma\ , \\
V_0 &= -\frac{1}{2}\bar{\mu}^{2}f_{\pi}^{2} + \frac{1}{4}\lambda_4 f_{\pi}^4 -\frac{1}{6}\lambda_6f_{\pi}^6 - m_{\pi}^{2} f_{\pi}^2\ .
\end{align}
Here, $i = +,-$ denotes the parity of nucleons, $f_{\pi}=92.4$ MeV represents the pion decay constant, and $E_{\mathbf{p}}^i = \sqrt{\mathbf{p}^2 + m_i^2}$ represents the energy of nucleons with mass $m_i$ and momentum $\mathbf{p}$. The parameters $\bar{\mu}^2, \lambda_4$ and $\lambda_6$ in the potential are constants which will be determined later. In conventional models, heavier degrees of freedom enter the theory when the chemical potential surpasses their mass threshold. However, the quarkyonic description modifies this picture. After quarks saturate and occupy the low-energy states, the emergence of excited states is suppressed due to the Pauli blocking of quarks. This quark  saturation shifts the onset of heavier degrees of freedom to higher chemical potentials.
Consequently, in this framework, the negative parity state $N(1535)$ enters the matter at a chemical potential higher than its mass. Moreover, at finite $N_c$, the quarkyonic picture is valid only within a limited window of quark chemical potential: $\Lambda_{\text{QCD}} < \mu_q < \sqrt{N_c}\Lambda_{\text{QCD}}$ \cite{McLerran:2007qj,McLerran:2008ua,Hidaka:2008yy,McLerran:2018hbz}. The upper bound of this range ($\mu_B = 3\mu_q$) is close to the mass of $N(1535)$. Given these considerations, in our study, we do not include the negative parity state in the density range that we are interested in. Confining forces remain only near the Fermi surface and nucleons appear in this momentum shell which is defined as\cite{McLerran:2018hbz}
\begin{align}
\Delta  = \frac{\Lambda^3_{{\rm QCD}}}{k^2_{FB}},
\label{eq_shell}
\end{align}
where $k_{FB}$ corresponds to the Fermi momentum of $N(939)$. It is important to note that this definition of momentum shell ensures that the nucleon density is approximately  $n_B \propto k_{FB}^2 \Delta \approx \Lambda_{{\rm QCD}}^3$.
Utilizing the thermodynamic relation $P = -\Omega$, we express the baryon part of the pressure $P_F$ over the Fermi momentum as
\begin{align}
    P_F =& P_{{ H}} + P_{{ Q}},\\
    P_{{ H}} =& 2  \sum_{\alpha=p, n}   \int^{k_{{ FB}}}_{N_c k_{{ FQ}}} \frac{\mathrm{d}^3 \mathbf{p}}{(2 \pi)^{3}}\left(\mu_{\alpha}^{*}-E_{{\bf p}}^{i}\right) ,\\
    P_{{ Q}} = & 4N_c \int_{0}^{k_{{ FQ}}} \frac{\mathrm{d}^3 \mathbf{q}}{(2 \pi)^{3}}(\mu_q^{*} - E_{{\bf q}}),
\label{eq_omega}
\end{align}
with
\begin{align}
    k_{FQ} &= \frac{k_{{ FB}}-\Delta}{N_c}\Theta(k_{FB} - \Delta),\\
    E_{{\bf q}} &= \sqrt{{\bf q}^2 + M_Q^2 }.
\end{align}
Here $M_Q$ is the constituent quark mass.
 We introduce a novel concept where the constituent quark also possesses an invariant mass component, analogous to the chiral invariant mass in the PDM. While we do not delve into the detailed origin of this invariant mass here, several scenarios are plausible.
The non-perturbative QCD vacuum is characterized by the presence of gluon condensates, which could contribute to an effective mass for quarks that persists even when chiral symmetry is restored. Also, QCD possesses rich topological structure, including instantons and other non-perturbative configurations, which could also generate an effective mass for quarks that is not directly tied to chiral symmetry breaking. This mass would be related to the fundamental structure of the QCD vacuum and would persist even as the chiral condensate diminishes. Furthermore, the precise mechanism of quark confinement in QCD remains an open question. Some models of confinement, such as those based on center vortices or dual superconductivity, suggest that the confining force could contribute to an effective quark mass\cite{Fischer:2004nq, Greensite:2016pfc} (See, eg. Ref.~\cite{Mazur:2020anw} for other possibility). This mass contribution would be largely independent of chiral symmetry breaking and could persist in regimes where chiral symmetry is restored. 
 
\begin{figure}[htbp]\centering
\includegraphics[width=1\hsize]{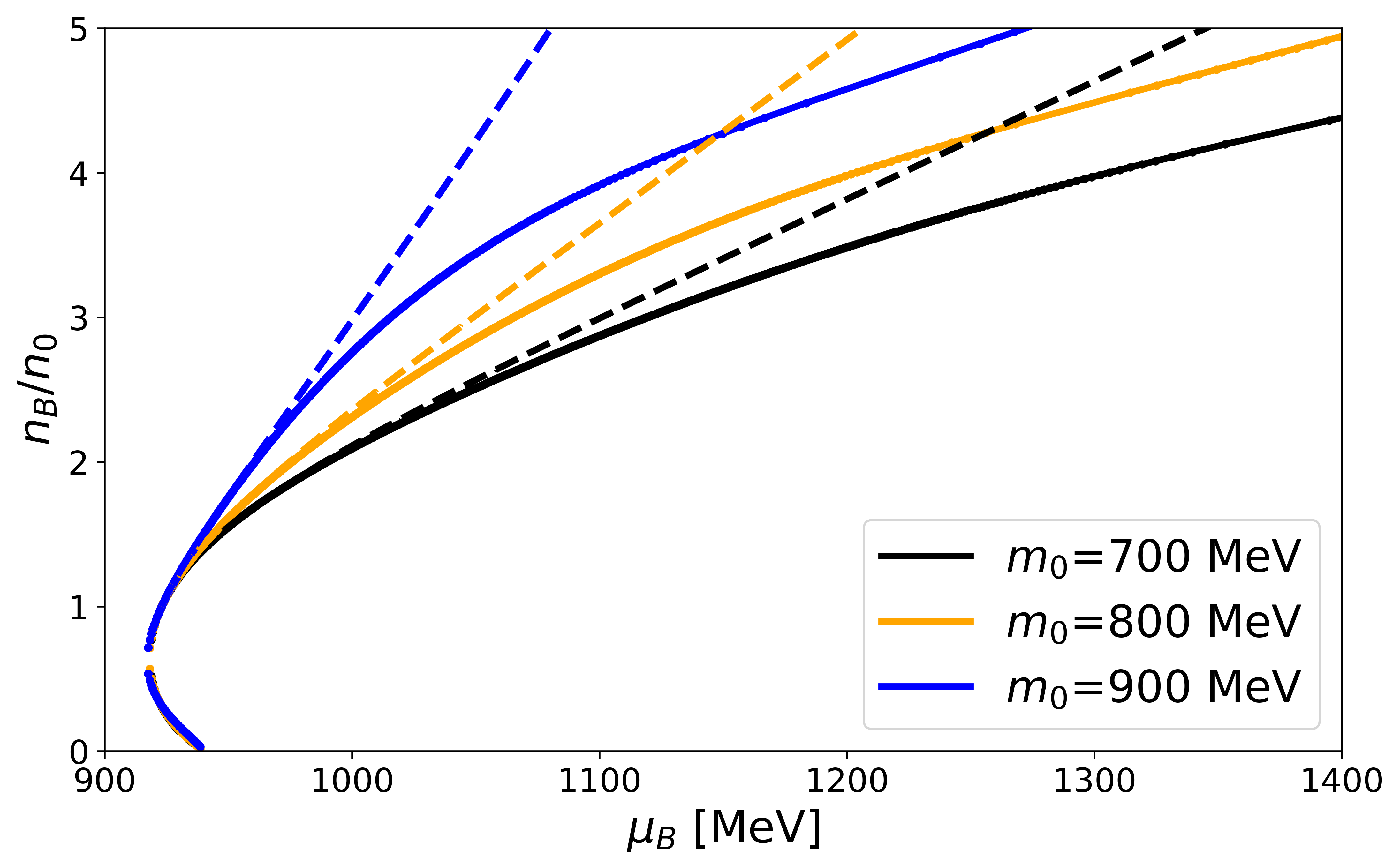}
\caption{Normalized baryon number density $n_B / n_0$ as a function of baryon chemical potential $\mu_B$ for different values of chiral invariant mass $m_0$. Solid curves represent the prediction of the current model, while dashed curves represent the results of pure PDM. }
\label{fig_nb_muB}
\end{figure}
 
The idea of quark-hadron continuity suggests a smooth transition between hadronic and quark degrees of freedom. An invariant mass component in constituent quarks could help explaining how some hadronic properties persist even in regimes where quark degrees of freedom become relevant. Models incorporating an invariant mass for baryons, such as the PDM, have been successful in describing various aspects of nuclear physics and neutron star properties. Extending this concept to constituent quarks provides a natural way to connect hadronic and quark-level descriptions. This invariant mass component in the constituent quark has implications for both the hadron mass spectrum and the EOS in our research. For simplicity, we first define the constituent quark mass and the nucleon masses as follows:
\begin{align}
    M_Q =& \frac{m_+}{3},\\
    m_\pm =& \sqrt{m_0^2 + \left(\frac{g_1 + g_2}{2}\right)^2} \mp \frac{g_1 - g_2}{2}\sigma.
\end{align}
Here, $g_1$ and $g_2$ are coupling constants determined by the vacuum values of $m_\pm$ and $m_0$ which represents the chiral invariant mass in the PDM. As density increases, chiral symmetry is gradually restored, leading the mean field $\sigma$ to approach zero. Consequently, the masses of positive and negative parity nucleons degenerate to $m_0$. This mechanism establishes a duality relation through the invariant mass.
\begin{figure*}[tbp]
\begin{center}
\includegraphics[width=17.5cm]{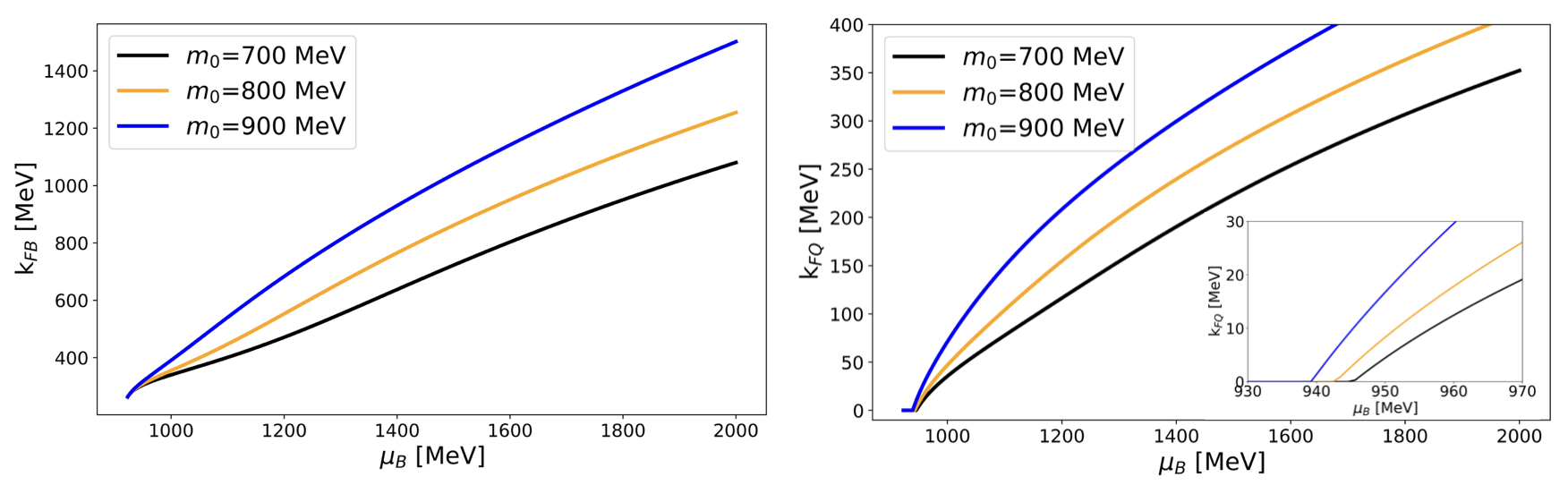}
\caption{Fermi momentum $k_{FB}$ (left panel) and $k_{FQ}$ (right panel) as a function of baryon chemical potential $\mu_B$ for several choices of chiral invariant mass $m_0$.  
}
\label{fig_k_muB}
\end{center}
\end{figure*}
\begin{figure}[htbp]\centering
\includegraphics[width=1\hsize]{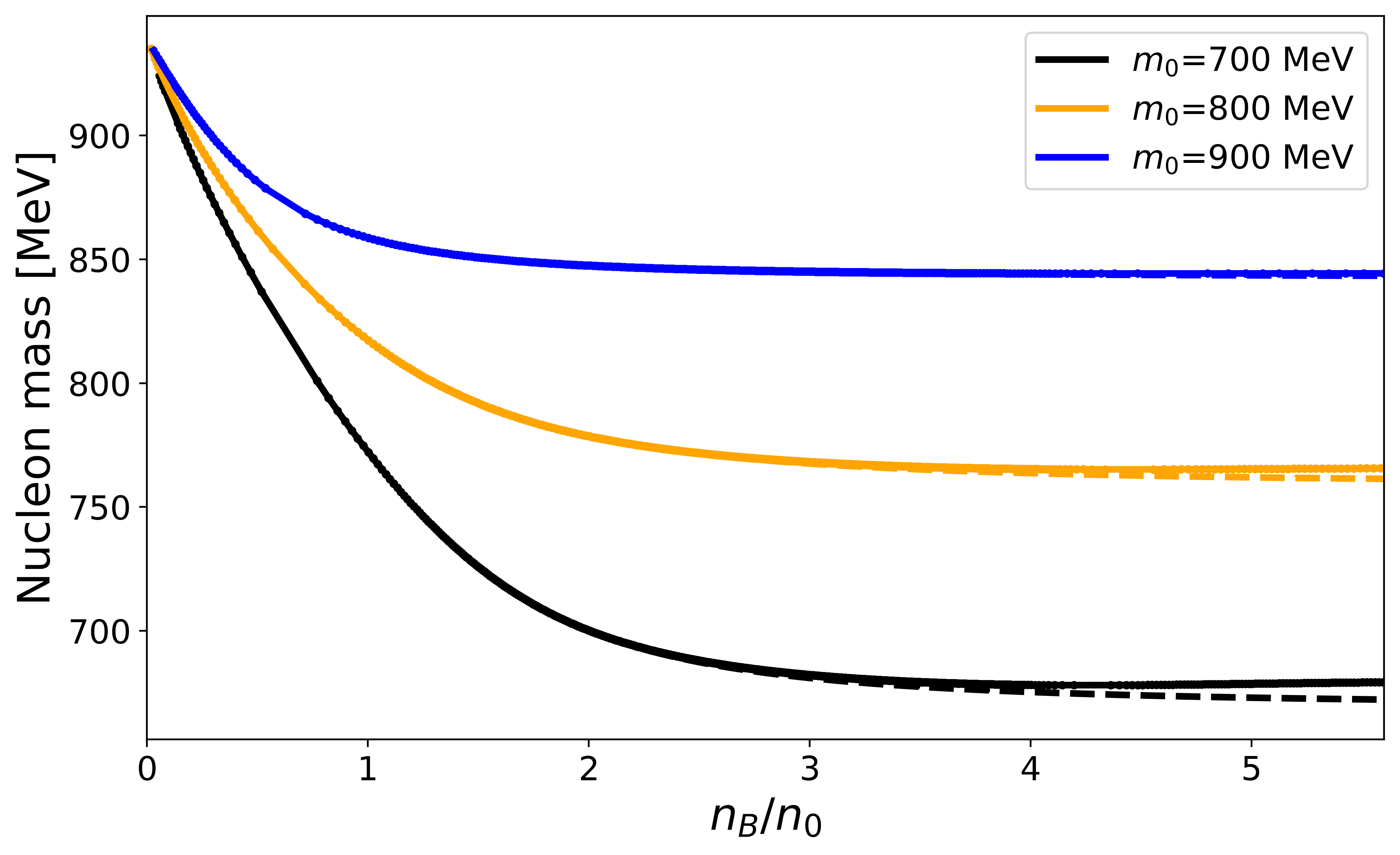}
\caption{Density dependence of the mass of $N(939)$ for  $m_0= 600, 700, 800$ MeV. }
\label{fig_mass_nb}
\end{figure}
From the familiar thermodynamic relations $n_B = \partial P / \partial \mu_B$, we calculate the baryon number density as  
\begin{align}
    n_{B}   = \frac{1}{3\pi^2}\sum\left[k_{FB}^3 - (N_c k_{FQ})^3\right] + \frac{2k_{FQ}^3}{3\pi^2}.
\label{eq_numberdensity}
\end{align}
The behavior of this system varies with density. At low densities, where $k_{FB}$ is small and the momentum shell $\Delta$ exceeds $k_{FB}$, we find $k_{FQ} = 0$. This phase corresponds to normal hadronic matter, where quark degrees of freedom do not contribute. As density increases and $k_{FB}$ surpasses $\Delta$, $k_{FQ}$ becomes non-zero, signaling the transition to the quarkyonic phase. From Eq.~(\ref{eq_numberdensity}), it is important to see that the contribution from the quarks relative to nucleons is suppressed by $1/N_c^3$. This formulation provides a unified description of the transition from hadronic to quarkyonic matter, incorporating both nucleon and quark degrees of freedom in a consistent framework.

In this study, we consider only the symmetric matter, treating up and down quarks equivalently. Here, the isospin density vanishes, resulting in a zero mean field $\rho$. At normal nuclear matter density $n_0 = 0.16$ fm$^{-3}$, we determine the parameters $\bar{\mu}^2$, $\lambda_4$, and $\lambda_6$ in the potential following the method outlined in Refs.\cite{Motohiro,Minamikawa:2020jfj,Gao:2024chh} for different values of $m_0$.
We then solve the gap equations:
\begin{equation}
\frac{\partial \Omega_{\text{PDM}}}{\partial \sigma} = 0, \quad \frac{\partial \Omega_{\text{PDM}}}{\partial \omega} = 0,
\label{eq_gap}
\end{equation}
to obtain the mean field values $\sigma$ and $\omega$ in finite density. We note that there is a relation between the choices of the chiral invariant mass and the stiffness of the EOS\cite{PhysRevC.100.025205,Minamikawa:2020jfj}. For a  smaller $m_0$, we need a larger scalar coupling to account for the nucleon mass, while it in turn demands a larger $\omega$ coupling due to the equilibrium state at the saturation density. As the density increases with the chiral restoration, the $\omega$ contributions become dominant, and then the EOSs for smaller $m_0$ become stiffer.

\section{Numerical results}
\label{sec_results}
After obtaining the mean filed $\sigma$ and $\omega$, we then calculate relevant physical quantities in this section.

In Fig.~\ref{fig_nb_muB}, we show the normalized baryon number density $n_B/n_0$ as a function of $\mu_B$ for several choices of chiral invariant masses $m_0$. Solid curves represent the quarkyonic description, while dashed curves represent the results in the ordinary PDM.
Previous studies\cite{PhysRevC.100.025205,Minamikawa:2020jfj} have demonstrated that larger $m_0$ typically results in softer EOS and larger baryon number density at a given baryon chemical potential as the dashed curves show in Fig.~\ref{fig_nb_muB}.

The differences appear as the system transitions into the quarkyonic phase, signaled by the emergence of a non-zero quark Fermi momentum $k_{FQ}$ (as shown in the right panel of Fig.~\ref{fig_k_muB}), the baryon number density in the present model is suppressed compared with the baryon number density of pure baryonic matter $n_B = \sum k^3_{FB} / 3\pi^2$  as  explained in Eq.~(\ref{eq_numberdensity}). Figure~\ref{fig_k_muB} illustrates that for a given $\mu_B$, both Fermi momenta $k_{FB}$ and $k_{FQ}$ increase with $m_0$. This phenomenon is a direct consequence of the larger baryon number density associated with larger $m_0$ at fixed $\mu_B$, which in turn increases the Fermi momenta as per Eq.~(\ref{eq_numberdensity}).

Furthermore, for larger values of $m_0$,  the quarkyonic matter appears in the lower $\mu_B$ region. This earlier transition can be attributed to the behavior of the momentum shell width $\Delta$ as in Eq.~(\ref{eq_shell}). For larger $m_0$, $\Delta$ decreases with increasing density more rapidly,  allowing quark degrees of freedom to become relevant at lower densities. This mechanism, relates to the larger Fermi momenta associated with larger $m_0$, facilitates the earlier appearance of quarkyonic matter in systems with higher chiral invariant masses.

In Fig.~\ref{fig_mass_nb}, we also show the density dependence of the mass of the ground state nucleon $N(939)$  for different values of $m_0$. The solid curves represent the nucleon mass in the quarkyonic description and the dashed curves show the density dependence of the nucleon mass in the PDM.
In the quarkyonic description, we observe a slight increase in the nucleon mass compared to the PDM.   However, this mass increment, while noticeable, is relatively small and does not affect the low-energy nucleon mass spectrum.

\begin{figure}[htbp]\centering
\includegraphics[width=1\hsize]{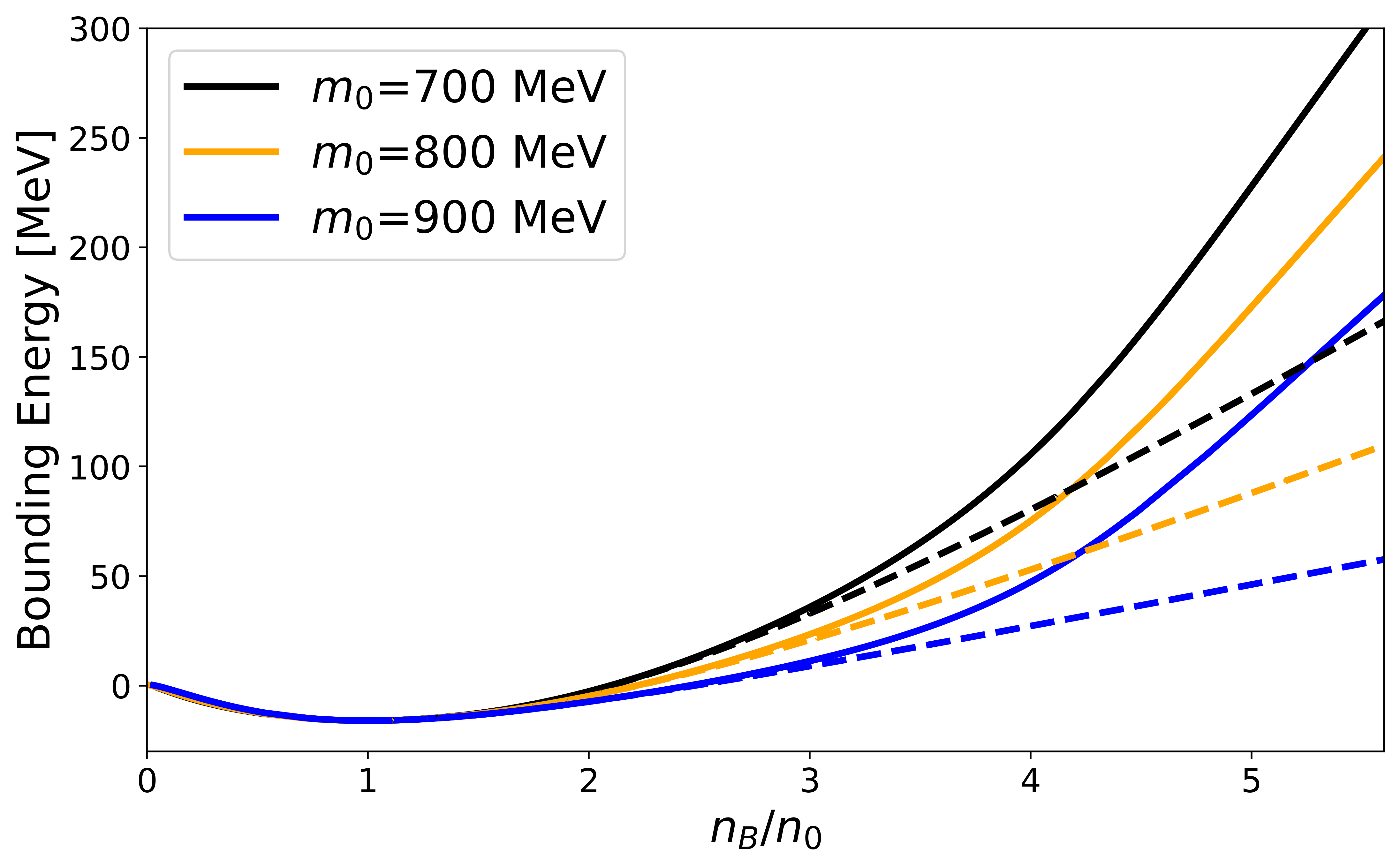}
\includegraphics[width=1\hsize]{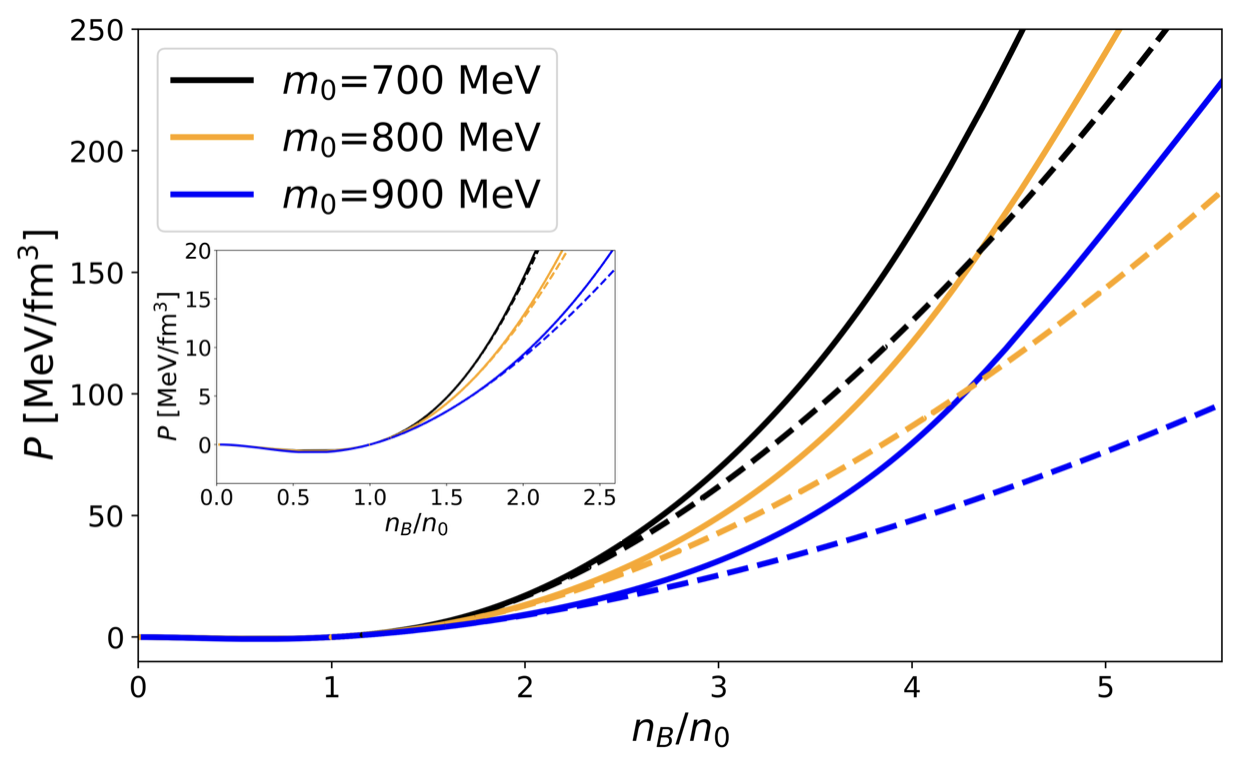}
\caption{Bounding energy $E/A - m_+$ (upper panel) and the pressure $P = -\Omega_{{\rm PDM}}$ (lower panel) as the function of normalized baryon number density $n_B / n_0$. The solid curves are for quarkyonic description and the dashed curves represent the results for the ordinary PDM.  }
\label{fig_eos}
\end{figure}

\begin{figure}[htbp]\centering
\includegraphics[width=1\hsize]{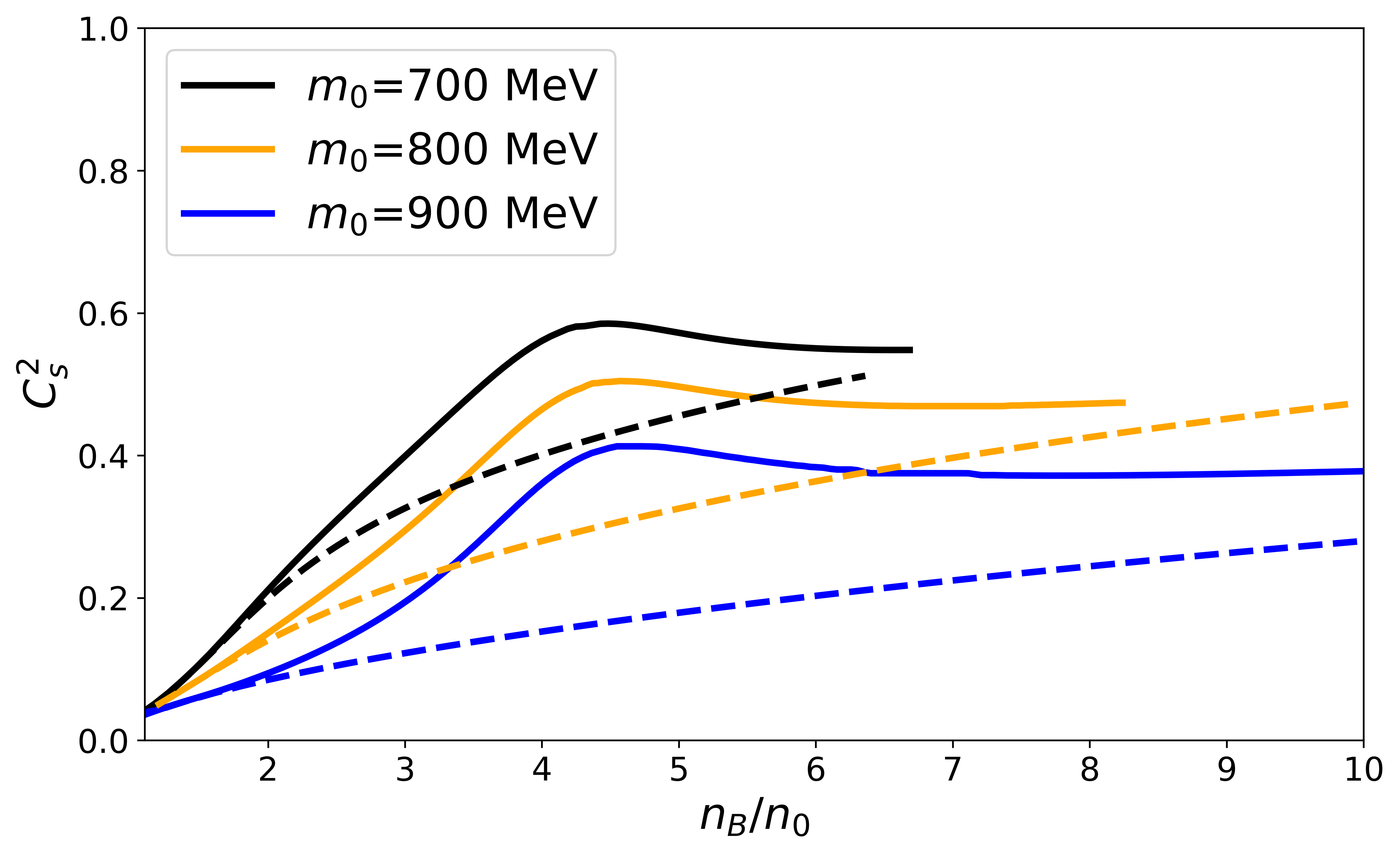}
\caption{Density dependence of the sound velocity $c_s^2 = {\rm d} p / {\rm d} \varepsilon$ for quarkyonic description (solid curves) and ordinary PDM (dashed curves) for several choices of $m_0$. }
\label{fig_soundvelocity}
\end{figure}

We present the results of bounding energy $E/A - m_+$ and pressure $P$ in Fig.~\ref{fig_eos}. The solid curves represent the quarkyonic description, while the dashed curves represent the results for the ordinary PDM. 
A significant observation is the stiffening of the EOS upon entering the quarkyonic phase  which occurs at approximately $1.4n_0$. This behavior can be directly attributed to the quark contribution to the pressure as 
\begin{align}
&4N_c \int_{0}^{k_{{ FQ}}} \frac{\mathrm{d}^3 \mathbf{q}}{(2 \pi)^{3}}(\mu_q^{*} - E_{{\bf q}}) \nonumber\\
&=   4N_c^4\int_0^{N_c k_{FQ}}\frac{\mathrm{d}^3 \mathbf{q^{\prime}}}{(2 \pi)^{3}}  \left(\mu_q^{*} - N_c\sqrt{( q^{\prime})^2 + (\frac{M_Q}{N_c})^2}\right).
\label{eq_requark}
\end{align}
Comparing this with Eq.~(\ref{eq_omega}), we see that the quark contribution relative to nucleons is enhanced by a factor of approximately $N_c^3$. This enhancement leads to a rapid increase of the sound velocity as shown in Fig.~\ref{fig_soundvelocity}, leading to a non-trivial peak structure in the intermediate density region.  
In ordinary hadron models, the sound velocity typically  monotonically increases  with density, different from the quarkyonic description.

\section{Invariant mass in constituent quark}\label{sec_constituent}
In this section, we examine the impact of including an invariant mass component in the constituent quark model. We propose a simple parametrization:
\begin{align}
M_{Q} &= m_+ / w(\sigma),\\
w(\sigma) &= w_0 - (w_0 -3)\frac{\sigma}{f_{\pi}}.
\end{align}
where $\omega_0$ is a constant parameter effectively help us to adjust the constituent quark mass after chiral symmetry restoration. This formulation offers a smooth transition between different regimes of chiral symmetry breaking. In the vacuum state, where $\sigma = f_{\pi}$, we recover $M_Q = m_+ / 3$, aligning with the conventional constituent quark model. As the chiral symmetry becomes restored ($\sigma \rightarrow 0$), the quark mass approaches to $M_Q = m_0 / w_0$, with the constraint $w_0 \geq 3$.
When $w_0 = 3$, the quark mass in the chirally restored phase derives entirely from the chiral invariant mass $m_0$, suggesting a direct connection between hadronic and quark properties. For $w_0 > 3$, quarks retain only a fraction of the nucleon's chiral invariant mass, potentially indicating additional mass-reduction mechanisms at play in dense matter. For very large value of $w_0$, the constituent quark mass for $\sigma \rightarrow 0$ is very small compared with 1/3 of the mass of nucleon, $M_{Q} \rightarrow m_+ / \omega_0 \ll m_+ / 3$.

\begin{figure}[htbp]\centering
\includegraphics[width=1\hsize]{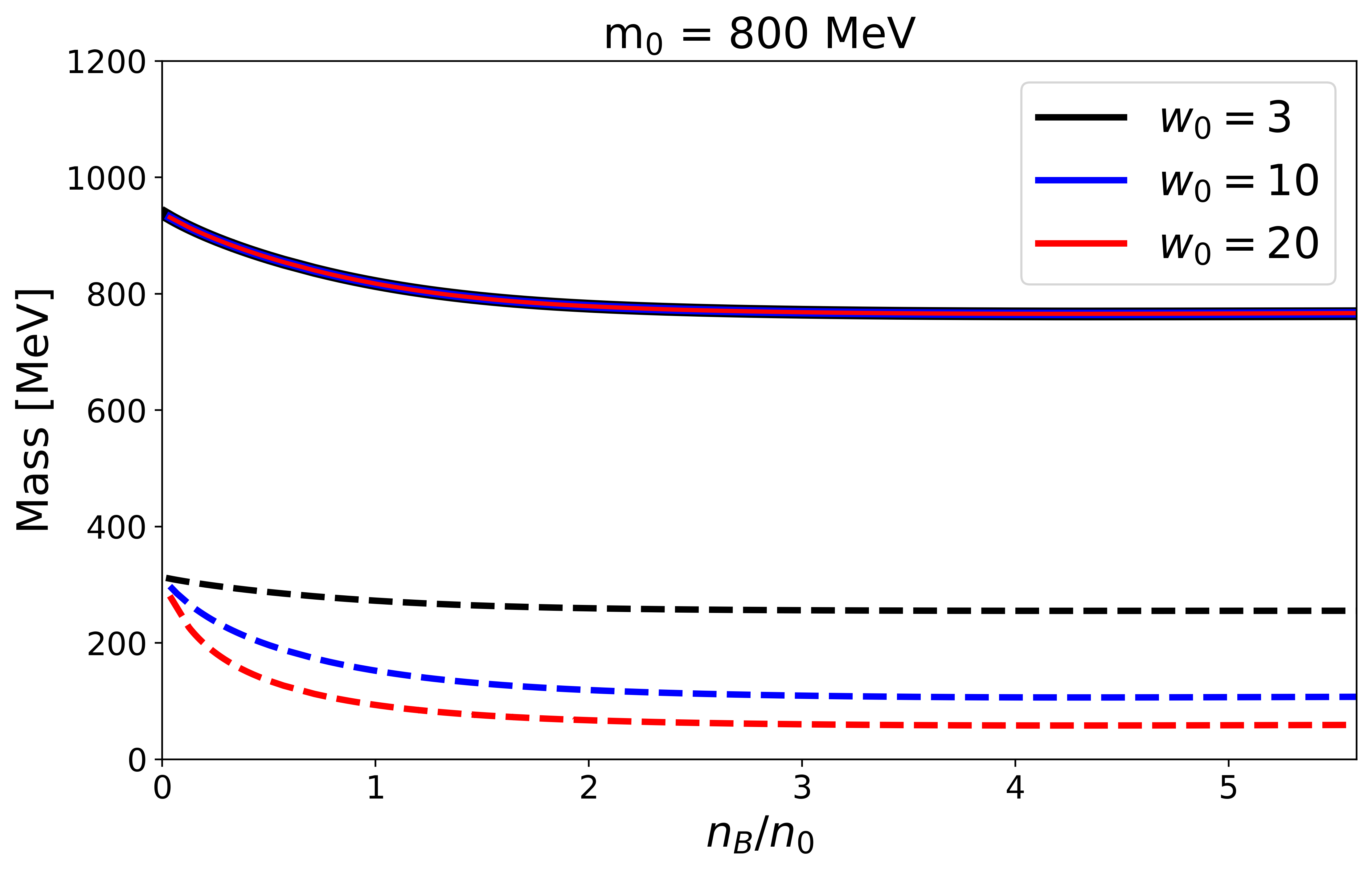}
\includegraphics[width=1\hsize]{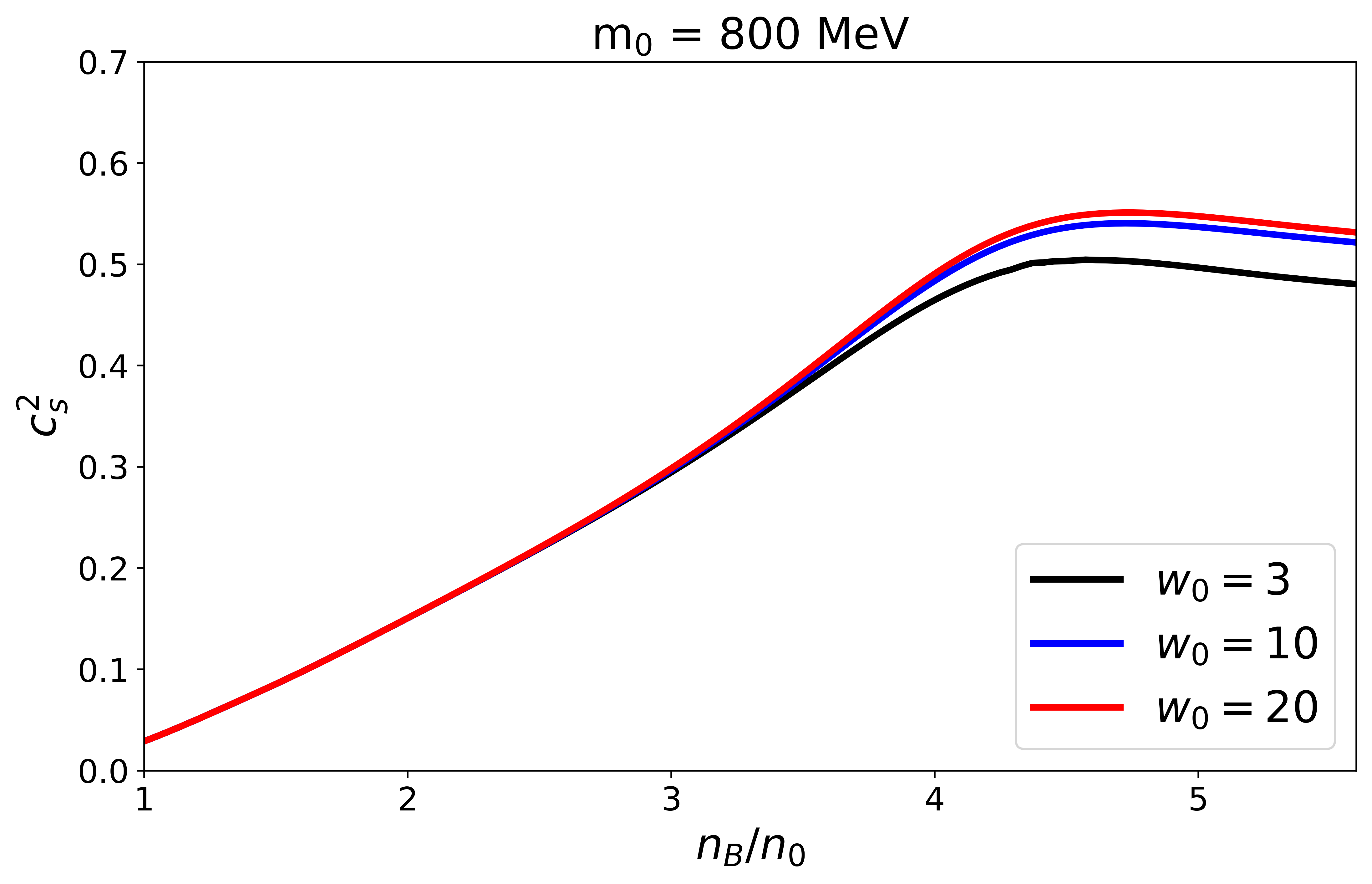}
\caption{ Upper panel: Density dependence of the nucleon mass (solid curve) and constituent quark mass (dashed curves) for several choices of $\omega_0$. Lower panel: Density dependence of the sound velocity for several choices of $\omega_0$.}
\label{fig_part2}
\end{figure}

Figure~\ref{fig_part2} illustrates the effects of varying $w_0$ on the nucleon mass, constituent quark mass, and sound velocity, with $m_0$ fixed at 800 MeV. The results reveal intricate relationships between quark substructure and macroscopic properties of dense matter. The upper panel demonstrates that changes in the invariant mass component of the constituent quark have a relatively modest impact on the nucleon mass. This behavior can be understood by examining the structure of the model. The nucleon mass is primarily determined by two factors: the chiral invariant mass $m_0$, which is independent of $w_0$, and the chiral condensate $\sigma$. While $w_0$ does not directly influence $m_0$, it affects the constituent quark mass, which in turn modifies the thermodynamic potential of the system. This modification feeds back into the gap equations that determine $\sigma$, creating an indirect link between $w_0$ and the nucleon mass. The observed moderate sensitivity of the nucleon mass to $w_0$ suggests a delicate balance in the model. This balance indicates that the chiral properties of the system remain relatively stable despite changes in the constituent quark mass structure. Consequently, we can infer that the presence of an invariant mass component in the constituent quark model do not significantly influence the low-density hadronic properties, such as decay widths or the hadron mass spectrum.

The lower panel in Fig.~\ref{fig_part2}, however, reveals the variations in the sound velocity  as $w_0$ changes. Notably, a smaller invariant mass component in the constituent quark (corresponding to larger $w_0$) leads to a larger values in the sound velocity. This behavior can be understood by drawing similarly with the original PDM analysis. In the PDM framework, a larger invariant mass component results in a weaker Yukawa coupling strength. Analogously, in our current model, when the invariant mass in the constituent quark becomes larger (smaller $w_0$), the Yukawa interaction of $\sigma$ to the constituent quark also  becomes weaker. This reduced interaction strength manifests as a smaller maximum value in the sound velocity.

\section{Summary and discussion}\label{sec_summary}
In this work, we have presented a novel approach to describe dense nuclear matter by integrating the quarkyonic matter framework with the PDM. This unified approach offers several key advancements in our understanding of the transition from hadronic to quark degrees of freedom in dense environments.

Our model introduces a chiral invariant mass component for both baryons and quarks, allowing for a more sophisticated treatment of chiral symmetry in dense matter. Unlike previous quarkyonic models that assumed constant particle masses, our approach permits both nucleon and constituent quark masses to decrease with increasing density. This feature more accurately reflects the expected behavior of nucleon and constituent quark in dense matter and provides a more realistic description of the system's evolution. Our model also incorporates interactions between nucleons, crucial for accurately describing the equation of state beyond saturation densities. The gradual restoration of chiral symmetry with increasing density is naturally incorporated, as evidenced by the behavior of nucleon and quark masses.

Our results demonstrate several interesting features. The equation of state exhibits a stiffening upon entering the quarkyonic phase, which we can now understand microscopically as an interplay between chiral symmetry restoration and the emergence of quark degrees of freedom. The sound velocity shows non-monotonic behavior, with a rapid increase at the onset of the quarkyonic phase followed by a more complex evolution at higher densities. 
The introduction of an invariant mass component in the constituent quark model, parametrized as $M_Q = m_+ / w(\sigma)$ where $w(\sigma) = w_0 - (w_0 - 3)\sigma/f_\pi$, affects the values of the sound velocity. Notably, a smaller invariant mass component in the constituent quark (corresponding to larger $w_0$) leads to a larger values of the sound velocity. This behavior can be understood by drawing parallels with the original PDM analysis, where a larger invariant mass component results in weaker Yukawa interactions. Our findings highlight the importance of quark substructure in determining the macroscopic properties of dense matter.
These results represent a significant step forward in the theoretical description of dense nuclear matter, bridging the gap between hadronic and quark degrees of freedom while respecting fundamental symmetries of QCD. The model's ability for smooth transition between hadronic and quark regimes, while incorporating chiral symmetry restoration effects, offers a more comprehensive picture of matter under extreme conditions.

The next crucial step of this research is to extend our model to neutron star matter, incorporating beta-equilibrium and charge neutrality conditions. This extension will enable direct comparisons with recent neutron star observations, including mass-radius measurements. Such comparisons will provide stringent tests for our model and potentially offer new insights into the composition and structure of neutron star cores.
Furthermore, exploring the implications of our model for other neutron star properties, such as cooling rates, glitch phenomena, and tidal deformabilities, could provide additional observational signatures of quarkyonic matter in neutron stars. These studies may help distinguishing different models of dense matter and shed light on the existence and properties of exotic phases in neutron star interiors.

In conclusion, our work represents a significant advancement in the theoretical description of dense nuclear matter. The application of this model to neutron star matter and subsequent comparisons with observational data will be crucial in validating its predictions and furthering our understanding of matter under extreme conditions. This research not only contributes to our fundamental understanding of nuclear physics but also has far-reaching implications for astrophysics and the study of compact objects in the universe.

\medskip
\acknowledgments
The authors acknowledge helpful conversations with Toru Kojo, Daiki Suenaga, Yuki Fujimoto and Larry D. McLerran. This work is supported in part by JSPS KAKENHI Grant Nos.~20K03927, 23H05439, 24K07045 and JST SPRING, Grant No. JPMJSP2125.  B.G. would like to take this opportunity to thank the
“Interdisciplinary Frontier Next-Generation Researcher Program of the Tokai
Higher Education and Research System.”

\bibliography{ref_quarkyonic}

\begin{thebibliography}{70}%
\makeatletter
\providecommand \@ifxundefined [1]{%
 \@ifx{#1\undefined}
}%
\providecommand \@ifnum [1]{%
 \ifnum #1\expandafter \@firstoftwo
 \else \expandafter \@secondoftwo
 \fi
}%
\providecommand \@ifx [1]{%
 \ifx #1\expandafter \@firstoftwo
 \else \expandafter \@secondoftwo
 \fi
}%
\providecommand \natexlab [1]{#1}%
\providecommand \enquote  [1]{``#1''}%
\providecommand \bibnamefont  [1]{#1}%
\providecommand \bibfnamefont [1]{#1}%
\providecommand \citenamefont [1]{#1}%
\providecommand \href@noop [0]{\@secondoftwo}%
\providecommand \href [0]{\begingroup \@sanitize@url \@href}%
\providecommand \@href[1]{\@@startlink{#1}\@@href}%
\providecommand \@@href[1]{\endgroup#1\@@endlink}%
\providecommand \@sanitize@url [0]{\catcode `\\12\catcode `\$12\catcode
  `\&12\catcode `\#12\catcode `\^12\catcode `\_12\catcode `\%12\relax}%
\providecommand \@@startlink[1]{}%
\providecommand \@@endlink[0]{}%
\providecommand \url  [0]{\begingroup\@sanitize@url \@url }%
\providecommand \@url [1]{\endgroup\@href {#1}{\urlprefix }}%
\providecommand \urlprefix  [0]{URL }%
\providecommand \Eprint [0]{\href }%
\providecommand \doibase [0]{http://dx.doi.org/}%
\providecommand \selectlanguage [0]{\@gobble}%
\providecommand \bibinfo  [0]{\@secondoftwo}%
\providecommand \bibfield  [0]{\@secondoftwo}%
\providecommand \translation [1]{[#1]}%
\providecommand \BibitemOpen [0]{}%
\providecommand \bibitemStop [0]{}%
\providecommand \bibitemNoStop [0]{.\EOS\space}%
\providecommand \EOS [0]{\spacefactor3000\relax}%
\providecommand \BibitemShut  [1]{\csname bibitem#1\endcsname}%
\let\auto@bib@innerbib\@empty
\bibitem [{\citenamefont {Fonseca}\ \emph {et~al.}(2016)\citenamefont {Fonseca}
  \emph {et~al.}}]{Fonseca:2016tux}%
  \BibitemOpen
  \bibfield  {author} {\bibinfo {author} {\bibfnamefont {E.}~\bibnamefont
  {Fonseca}} \emph {et~al.},\ }\href {\doibase 10.3847/0004-637X/832/2/167}
  {\bibfield  {journal} {\bibinfo  {journal} {Astrophys. J.}\ }\textbf
  {\bibinfo {volume} {832}},\ \bibinfo {pages} {167} (\bibinfo {year}
  {2016})},\ \Eprint {http://arxiv.org/abs/1603.00545} {arXiv:1603.00545
  [astro-ph.HE]} \BibitemShut {NoStop}%
\bibitem [{\citenamefont {Abbott}\ \emph
  {et~al.}(2017{\natexlab{a}})\citenamefont {Abbott} \emph
  {et~al.}}]{LIGOScientific:2017vwq}%
  \BibitemOpen
  \bibfield  {author} {\bibinfo {author} {\bibfnamefont {B.~P.}\ \bibnamefont
  {Abbott}} \emph {et~al.} (\bibinfo {collaboration} {LIGO Scientific,
  Virgo}),\ }\href {\doibase 10.1103/PhysRevLett.119.161101} {\bibfield
  {journal} {\bibinfo  {journal} {Phys. Rev. Lett.}\ }\textbf {\bibinfo
  {volume} {119}},\ \bibinfo {pages} {161101} (\bibinfo {year}
  {2017}{\natexlab{a}})},\ \Eprint {http://arxiv.org/abs/1710.05832}
  {arXiv:1710.05832 [gr-qc]} \BibitemShut {NoStop}%
\bibitem [{\citenamefont {Abbott}\ \emph
  {et~al.}(2017{\natexlab{b}})\citenamefont {Abbott} \emph
  {et~al.}}]{LIGOScientific:2017ync}%
  \BibitemOpen
  \bibfield  {author} {\bibinfo {author} {\bibfnamefont {B.~P.}\ \bibnamefont
  {Abbott}} \emph {et~al.} (\bibinfo {collaboration} {LIGO Scientific, Virgo,
  Fermi GBM, INTEGRAL, IceCube, AstroSat Cadmium Zinc Telluride Imager Team,
  IPN, Insight-Hxmt, ANTARES, Swift, AGILE Team, 1M2H Team, Dark Energy Camera
  GW-EM, DES, DLT40, GRAWITA, Fermi-LAT, ATCA, ASKAP, Las Cumbres Observatory
  Group, OzGrav, DWF (Deeper Wider Faster Program), AST3, CAASTRO, VINROUGE,
  MASTER, J-GEM, GROWTH, JAGWAR, CaltechNRAO, TTU-NRAO, NuSTAR, Pan-STARRS,
  MAXI Team, TZAC Consortium, KU, Nordic Optical Telescope, ePESSTO, GROND,
  Texas Tech University, SALT Group, TOROS, BOOTES, MWA, CALET, IKI-GW
  Follow-up, H.E.S.S., LOFAR, LWA, HAWC, Pierre Auger, ALMA, Euro VLBI Team, Pi
  of Sky, Chandra Team at McGill University, DFN, ATLAS Telescopes, High Time
  Resolution Universe Survey, RIMAS, RATIR, SKA South Africa/MeerKAT}),\ }\href
  {\doibase 10.3847/2041-8213/aa91c9} {\bibfield  {journal} {\bibinfo
  {journal} {Astrophys. J. Lett.}\ }\textbf {\bibinfo {volume} {848}},\
  \bibinfo {pages} {L12} (\bibinfo {year} {2017}{\natexlab{b}})},\ \Eprint
  {http://arxiv.org/abs/1710.05833} {arXiv:1710.05833 [astro-ph.HE]}
  \BibitemShut {NoStop}%
\bibitem [{\citenamefont {Miller}\ \emph {et~al.}(2021)\citenamefont {Miller}
  \emph {et~al.}}]{Miller:2021qha}%
  \BibitemOpen
  \bibfield  {author} {\bibinfo {author} {\bibfnamefont {M.~C.}\ \bibnamefont
  {Miller}} \emph {et~al.},\ }\href {\doibase 10.3847/2041-8213/ac089b}
  {\bibfield  {journal} {\bibinfo  {journal} {Astrophys. J. Lett.}\ }\textbf
  {\bibinfo {volume} {918}},\ \bibinfo {pages} {L28} (\bibinfo {year}
  {2021})},\ \Eprint {http://arxiv.org/abs/2105.06979} {arXiv:2105.06979
  [astro-ph.HE]} \BibitemShut {NoStop}%
\bibitem [{\citenamefont {Riley}\ \emph {et~al.}(2021)\citenamefont {Riley},
  \citenamefont {Watts}, \citenamefont {Ray}, \citenamefont {Bogdanov},
  \citenamefont {Guillot}, \citenamefont {Morsink}, \citenamefont {Bilous},
  \citenamefont {Arzoumanian}, \citenamefont {Choudhury}, \citenamefont
  {Deneva}, \citenamefont {Gendreau}, \citenamefont {Harding}, \citenamefont
  {Ho}, \citenamefont {Lattimer}, \citenamefont {Loewenstein}, \citenamefont
  {Ludlam}, \citenamefont {Markwardt}, \citenamefont {Okajima}, \citenamefont
  {Prescod-Weinstein}, \citenamefont {Remillard}, \citenamefont {Wolff},
  \citenamefont {Fonseca}, \citenamefont {Cromartie}, \citenamefont {Kerr},
  \citenamefont {Pennucci}, \citenamefont {Parthasarathy}, \citenamefont
  {Ransom}, \citenamefont {Stairs}, \citenamefont {Guillemot},\ and\
  \citenamefont {Cognard}}]{Riley:2021pdl}%
  \BibitemOpen
  \bibfield  {author} {\bibinfo {author} {\bibfnamefont {T.~E.}\ \bibnamefont
  {Riley}}, \bibinfo {author} {\bibfnamefont {A.~L.}\ \bibnamefont {Watts}},
  \bibinfo {author} {\bibfnamefont {P.~S.}\ \bibnamefont {Ray}}, \bibinfo
  {author} {\bibfnamefont {S.}~\bibnamefont {Bogdanov}}, \bibinfo {author}
  {\bibfnamefont {S.}~\bibnamefont {Guillot}}, \bibinfo {author} {\bibfnamefont
  {S.~M.}\ \bibnamefont {Morsink}}, \bibinfo {author} {\bibfnamefont {A.~V.}\
  \bibnamefont {Bilous}}, \bibinfo {author} {\bibfnamefont {Z.}~\bibnamefont
  {Arzoumanian}}, \bibinfo {author} {\bibfnamefont {D.}~\bibnamefont
  {Choudhury}}, \bibinfo {author} {\bibfnamefont {J.~S.}\ \bibnamefont
  {Deneva}}, \bibinfo {author} {\bibfnamefont {K.~C.}\ \bibnamefont
  {Gendreau}}, \bibinfo {author} {\bibfnamefont {A.~K.}\ \bibnamefont
  {Harding}}, \bibinfo {author} {\bibfnamefont {W.~C.~G.}\ \bibnamefont {Ho}},
  \bibinfo {author} {\bibfnamefont {J.~M.}\ \bibnamefont {Lattimer}}, \bibinfo
  {author} {\bibfnamefont {M.}~\bibnamefont {Loewenstein}}, \bibinfo {author}
  {\bibfnamefont {R.~M.}\ \bibnamefont {Ludlam}}, \bibinfo {author}
  {\bibfnamefont {C.~B.}\ \bibnamefont {Markwardt}}, \bibinfo {author}
  {\bibfnamefont {T.}~\bibnamefont {Okajima}}, \bibinfo {author} {\bibfnamefont
  {C.}~\bibnamefont {Prescod-Weinstein}}, \bibinfo {author} {\bibfnamefont
  {R.~A.}\ \bibnamefont {Remillard}}, \bibinfo {author} {\bibfnamefont {M.~T.}\
  \bibnamefont {Wolff}}, \bibinfo {author} {\bibfnamefont {E.}~\bibnamefont
  {Fonseca}}, \bibinfo {author} {\bibfnamefont {H.~T.}\ \bibnamefont
  {Cromartie}}, \bibinfo {author} {\bibfnamefont {M.}~\bibnamefont {Kerr}},
  \bibinfo {author} {\bibfnamefont {T.~T.}\ \bibnamefont {Pennucci}}, \bibinfo
  {author} {\bibfnamefont {A.}~\bibnamefont {Parthasarathy}}, \bibinfo {author}
  {\bibfnamefont {S.}~\bibnamefont {Ransom}}, \bibinfo {author} {\bibfnamefont
  {I.}~\bibnamefont {Stairs}}, \bibinfo {author} {\bibfnamefont
  {L.}~\bibnamefont {Guillemot}}, \ and\ \bibinfo {author} {\bibfnamefont
  {I.}~\bibnamefont {Cognard}},\ }\href {\doibase 10.3847/2041-8213/ac0a81}
  {\bibfield  {journal} {\bibinfo  {journal} {The Astrophysical Journal
  Letters}\ }\textbf {\bibinfo {volume} {918}},\ \bibinfo {pages} {L27}
  (\bibinfo {year} {2021})}\BibitemShut {NoStop}%
\bibitem [{\citenamefont {Fonseca}\ \emph {et~al.}(2021)\citenamefont {Fonseca}
  \emph {et~al.}}]{Fonseca:2021wxt}%
  \BibitemOpen
  \bibfield  {author} {\bibinfo {author} {\bibfnamefont {E.}~\bibnamefont
  {Fonseca}} \emph {et~al.},\ }\href {\doibase 10.3847/2041-8213/ac03b8}
  {\bibfield  {journal} {\bibinfo  {journal} {Astrophys. J. Lett.}\ }\textbf
  {\bibinfo {volume} {915}},\ \bibinfo {pages} {L12} (\bibinfo {year}
  {2021})},\ \Eprint {http://arxiv.org/abs/2104.00880} {arXiv:2104.00880
  [astro-ph.HE]} \BibitemShut {NoStop}%
\bibitem [{\citenamefont {Vinciguerra}\ \emph {et~al.}(2024)\citenamefont
  {Vinciguerra} \emph {et~al.}}]{Vinciguerra:2023qxq}%
  \BibitemOpen
  \bibfield  {author} {\bibinfo {author} {\bibfnamefont {S.}~\bibnamefont
  {Vinciguerra}} \emph {et~al.},\ }\href {\doibase 10.3847/1538-4357/acfb83}
  {\bibfield  {journal} {\bibinfo  {journal} {Astrophys. J.}\ }\textbf
  {\bibinfo {volume} {961}},\ \bibinfo {pages} {62} (\bibinfo {year} {2024})},\
  \Eprint {http://arxiv.org/abs/2308.09469} {arXiv:2308.09469 [astro-ph.HE]}
  \BibitemShut {NoStop}%
\bibitem [{\citenamefont {Lenzi}\ and\ \citenamefont
  {Lugones}(2012)}]{Lenzi:2012xz}%
  \BibitemOpen
  \bibfield  {author} {\bibinfo {author} {\bibfnamefont {C.~H.}\ \bibnamefont
  {Lenzi}}\ and\ \bibinfo {author} {\bibfnamefont {G.}~\bibnamefont
  {Lugones}},\ }\href {\doibase 10.1088/0004-637X/759/1/57} {\bibfield
  {journal} {\bibinfo  {journal} {Astrophys. J.}\ }\textbf {\bibinfo {volume}
  {759}},\ \bibinfo {pages} {57} (\bibinfo {year} {2012})},\ \Eprint
  {http://arxiv.org/abs/1206.4108} {arXiv:1206.4108 [astro-ph.SR]} \BibitemShut
  {NoStop}%
\bibitem [{\citenamefont {Benic}\ \emph {et~al.}(2015)\citenamefont {Benic},
  \citenamefont {Blaschke}, \citenamefont {Alvarez-Castillo}, \citenamefont
  {Fischer},\ and\ \citenamefont {Typel}}]{Benic:2014jia}%
  \BibitemOpen
  \bibfield  {author} {\bibinfo {author} {\bibfnamefont {S.}~\bibnamefont
  {Benic}}, \bibinfo {author} {\bibfnamefont {D.}~\bibnamefont {Blaschke}},
  \bibinfo {author} {\bibfnamefont {D.~E.}\ \bibnamefont {Alvarez-Castillo}},
  \bibinfo {author} {\bibfnamefont {T.}~\bibnamefont {Fischer}}, \ and\
  \bibinfo {author} {\bibfnamefont {S.}~\bibnamefont {Typel}},\ }\href
  {\doibase 10.1051/0004-6361/201425318} {\bibfield  {journal} {\bibinfo
  {journal} {Astron. Astrophys.}\ }\textbf {\bibinfo {volume} {577}},\ \bibinfo
  {pages} {A40} (\bibinfo {year} {2015})},\ \Eprint
  {http://arxiv.org/abs/1411.2856} {arXiv:1411.2856 [astro-ph.HE]} \BibitemShut
  {NoStop}%
\bibitem [{\citenamefont {Zdunik}\ and\ \citenamefont
  {Haensel}(2013)}]{Zdunik:2012dj}%
  \BibitemOpen
  \bibfield  {author} {\bibinfo {author} {\bibfnamefont {J.~L.}\ \bibnamefont
  {Zdunik}}\ and\ \bibinfo {author} {\bibfnamefont {P.}~\bibnamefont
  {Haensel}},\ }\href {\doibase 10.1051/0004-6361/201220697} {\bibfield
  {journal} {\bibinfo  {journal} {Astron. Astrophys.}\ }\textbf {\bibinfo
  {volume} {551}},\ \bibinfo {pages} {A61} (\bibinfo {year} {2013})},\ \Eprint
  {http://arxiv.org/abs/1211.1231} {arXiv:1211.1231 [astro-ph.SR]} \BibitemShut
  {NoStop}%
\bibitem [{\citenamefont {Alford}\ \emph {et~al.}(2013)\citenamefont {Alford},
  \citenamefont {Han},\ and\ \citenamefont {Prakash}}]{Alford:2013aca}%
  \BibitemOpen
  \bibfield  {author} {\bibinfo {author} {\bibfnamefont {M.~G.}\ \bibnamefont
  {Alford}}, \bibinfo {author} {\bibfnamefont {S.}~\bibnamefont {Han}}, \ and\
  \bibinfo {author} {\bibfnamefont {M.}~\bibnamefont {Prakash}},\ }\href
  {\doibase 10.1103/PhysRevD.88.083013} {\bibfield  {journal} {\bibinfo
  {journal} {Phys. Rev. D}\ }\textbf {\bibinfo {volume} {88}},\ \bibinfo
  {pages} {083013} (\bibinfo {year} {2013})},\ \Eprint
  {http://arxiv.org/abs/1302.4732} {arXiv:1302.4732 [astro-ph.SR]} \BibitemShut
  {NoStop}%
\bibitem [{\citenamefont {G\"artlein}\ \emph {et~al.}(2023)\citenamefont
  {G\"artlein}, \citenamefont {Ivanytskyi}, \citenamefont {Sagun},\ and\
  \citenamefont {Blaschke}}]{Gartlein:2023vif}%
  \BibitemOpen
  \bibfield  {author} {\bibinfo {author} {\bibfnamefont {C.}~\bibnamefont
  {G\"artlein}}, \bibinfo {author} {\bibfnamefont {O.}~\bibnamefont
  {Ivanytskyi}}, \bibinfo {author} {\bibfnamefont {V.}~\bibnamefont {Sagun}}, \
  and\ \bibinfo {author} {\bibfnamefont {D.}~\bibnamefont {Blaschke}},\ }\href
  {\doibase 10.1103/PhysRevD.108.114028} {\bibfield  {journal} {\bibinfo
  {journal} {Phys. Rev. D}\ }\textbf {\bibinfo {volume} {108}},\ \bibinfo
  {pages} {114028} (\bibinfo {year} {2023})},\ \Eprint
  {http://arxiv.org/abs/2301.10765} {arXiv:2301.10765 [nucl-th]} \BibitemShut
  {NoStop}%
\bibitem [{\citenamefont {Masuda}\ \emph {et~al.}(2013)\citenamefont {Masuda},
  \citenamefont {Hatsuda},\ and\ \citenamefont {Takatsuka}}]{Masuda:2012kf}%
  \BibitemOpen
  \bibfield  {author} {\bibinfo {author} {\bibfnamefont {K.}~\bibnamefont
  {Masuda}}, \bibinfo {author} {\bibfnamefont {T.}~\bibnamefont {Hatsuda}}, \
  and\ \bibinfo {author} {\bibfnamefont {T.}~\bibnamefont {Takatsuka}},\ }\href
  {\doibase 10.1088/0004-637X/764/1/12} {\bibfield  {journal} {\bibinfo
  {journal} {Astrophys. J.}\ }\textbf {\bibinfo {volume} {764}},\ \bibinfo
  {pages} {12} (\bibinfo {year} {2013})},\ \Eprint
  {http://arxiv.org/abs/1205.3621} {arXiv:1205.3621 [nucl-th]} \BibitemShut
  {NoStop}%
\bibitem [{\citenamefont {Kojo}\ \emph {et~al.}(2015)\citenamefont {Kojo},
  \citenamefont {Powell}, \citenamefont {Song},\ and\ \citenamefont
  {Baym}}]{Kojo:2014rca}%
  \BibitemOpen
  \bibfield  {author} {\bibinfo {author} {\bibfnamefont {T.}~\bibnamefont
  {Kojo}}, \bibinfo {author} {\bibfnamefont {P.~D.}\ \bibnamefont {Powell}},
  \bibinfo {author} {\bibfnamefont {Y.}~\bibnamefont {Song}}, \ and\ \bibinfo
  {author} {\bibfnamefont {G.}~\bibnamefont {Baym}},\ }\href {\doibase
  10.1103/PhysRevD.91.045003} {\bibfield  {journal} {\bibinfo  {journal} {Phys.
  Rev. D}\ }\textbf {\bibinfo {volume} {91}},\ \bibinfo {pages} {045003}
  (\bibinfo {year} {2015})},\ \Eprint {http://arxiv.org/abs/1412.1108}
  {arXiv:1412.1108 [hep-ph]} \BibitemShut {NoStop}%
\bibitem [{\citenamefont {Baym}\ \emph {et~al.}(2018)\citenamefont {Baym},
  \citenamefont {Hatsuda}, \citenamefont {Kojo}, \citenamefont {Powell},
  \citenamefont {Song},\ and\ \citenamefont {Takatsuka}}]{Baym:2017whm}%
  \BibitemOpen
  \bibfield  {author} {\bibinfo {author} {\bibfnamefont {G.}~\bibnamefont
  {Baym}}, \bibinfo {author} {\bibfnamefont {T.}~\bibnamefont {Hatsuda}},
  \bibinfo {author} {\bibfnamefont {T.}~\bibnamefont {Kojo}}, \bibinfo {author}
  {\bibfnamefont {P.~D.}\ \bibnamefont {Powell}}, \bibinfo {author}
  {\bibfnamefont {Y.}~\bibnamefont {Song}}, \ and\ \bibinfo {author}
  {\bibfnamefont {T.}~\bibnamefont {Takatsuka}},\ }\href {\doibase
  10.1088/1361-6633/aaae14} {\bibfield  {journal} {\bibinfo  {journal} {Rept.
  Prog. Phys.}\ }\textbf {\bibinfo {volume} {81}},\ \bibinfo {pages} {056902}
  (\bibinfo {year} {2018})},\ \Eprint {http://arxiv.org/abs/1707.04966}
  {arXiv:1707.04966 [astro-ph.HE]} \BibitemShut {NoStop}%
\bibitem [{\citenamefont {Baym}\ \emph {et~al.}(2019)\citenamefont {Baym},
  \citenamefont {Furusawa}, \citenamefont {Hatsuda}, \citenamefont {Kojo},\
  and\ \citenamefont {Togashi}}]{Baym:2019iky}%
  \BibitemOpen
  \bibfield  {author} {\bibinfo {author} {\bibfnamefont {G.}~\bibnamefont
  {Baym}}, \bibinfo {author} {\bibfnamefont {S.}~\bibnamefont {Furusawa}},
  \bibinfo {author} {\bibfnamefont {T.}~\bibnamefont {Hatsuda}}, \bibinfo
  {author} {\bibfnamefont {T.}~\bibnamefont {Kojo}}, \ and\ \bibinfo {author}
  {\bibfnamefont {H.}~\bibnamefont {Togashi}},\ }\href {\doibase
  10.3847/1538-4357/ab441e} {\bibfield  {journal} {\bibinfo  {journal}
  {Astrophys. J.}\ }\textbf {\bibinfo {volume} {885}},\ \bibinfo {pages} {42}
  (\bibinfo {year} {2019})},\ \Eprint {http://arxiv.org/abs/1903.08963}
  {arXiv:1903.08963 [astro-ph.HE]} \BibitemShut {NoStop}%
\bibitem [{\citenamefont {Kojo}\ \emph {et~al.}(2022)\citenamefont {Kojo},
  \citenamefont {Baym},\ and\ \citenamefont {Hatsuda}}]{Kojo:2021wax}%
  \BibitemOpen
  \bibfield  {author} {\bibinfo {author} {\bibfnamefont {T.}~\bibnamefont
  {Kojo}}, \bibinfo {author} {\bibfnamefont {G.}~\bibnamefont {Baym}}, \ and\
  \bibinfo {author} {\bibfnamefont {T.}~\bibnamefont {Hatsuda}},\ }\href
  {\doibase 10.3847/1538-4357/ac7876} {\bibfield  {journal} {\bibinfo
  {journal} {Astrophys. J.}\ }\textbf {\bibinfo {volume} {934}},\ \bibinfo
  {pages} {46} (\bibinfo {year} {2022})},\ \Eprint
  {http://arxiv.org/abs/2111.11919} {arXiv:2111.11919 [astro-ph.HE]}
  \BibitemShut {NoStop}%
\bibitem [{\citenamefont {Gao}\ \emph {et~al.}(2022)\citenamefont {Gao},
  \citenamefont {Minamikawa}, \citenamefont {Kojo},\ and\ \citenamefont
  {Harada}}]{Gao:2022klm}%
  \BibitemOpen
  \bibfield  {author} {\bibinfo {author} {\bibfnamefont {B.}~\bibnamefont
  {Gao}}, \bibinfo {author} {\bibfnamefont {T.}~\bibnamefont {Minamikawa}},
  \bibinfo {author} {\bibfnamefont {T.}~\bibnamefont {Kojo}}, \ and\ \bibinfo
  {author} {\bibfnamefont {M.}~\bibnamefont {Harada}},\ }\href {\doibase
  10.1103/PhysRevC.106.065205} {\bibfield  {journal} {\bibinfo  {journal}
  {Phys. Rev. C}\ }\textbf {\bibinfo {volume} {106}},\ \bibinfo {pages}
  {065205} (\bibinfo {year} {2022})},\ \Eprint
  {http://arxiv.org/abs/2207.05970} {arXiv:2207.05970 [nucl-th]} \BibitemShut
  {NoStop}%
\bibitem [{\citenamefont {Minamikawa}\ \emph
  {et~al.}(2023{\natexlab{a}})\citenamefont {Minamikawa}, \citenamefont {Gao},
  \citenamefont {Kojo},\ and\ \citenamefont {Harada}}]{Minamikawa:2023eky}%
  \BibitemOpen
  \bibfield  {author} {\bibinfo {author} {\bibfnamefont {T.}~\bibnamefont
  {Minamikawa}}, \bibinfo {author} {\bibfnamefont {B.}~\bibnamefont {Gao}},
  \bibinfo {author} {\bibfnamefont {T.}~\bibnamefont {Kojo}}, \ and\ \bibinfo
  {author} {\bibfnamefont {M.}~\bibnamefont {Harada}},\ }\href {\doibase
  10.3390/sym15030745} {\bibfield  {journal} {\bibinfo  {journal} {Symmetry}\
  }\textbf {\bibinfo {volume} {15}},\ \bibinfo {pages} {745} (\bibinfo {year}
  {2023}{\natexlab{a}})},\ \Eprint {http://arxiv.org/abs/2302.00825}
  {arXiv:2302.00825 [nucl-th]} \BibitemShut {NoStop}%
\bibitem [{\citenamefont {McLerran}\ and\ \citenamefont
  {Pisarski}(2007)}]{McLerran:2007qj}%
  \BibitemOpen
  \bibfield  {author} {\bibinfo {author} {\bibfnamefont {L.}~\bibnamefont
  {McLerran}}\ and\ \bibinfo {author} {\bibfnamefont {R.~D.}\ \bibnamefont
  {Pisarski}},\ }\href {\doibase 10.1016/j.nuclphysa.2007.08.013} {\bibfield
  {journal} {\bibinfo  {journal} {Nucl. Phys. A}\ }\textbf {\bibinfo {volume}
  {796}},\ \bibinfo {pages} {83} (\bibinfo {year} {2007})},\ \Eprint
  {http://arxiv.org/abs/0706.2191} {arXiv:0706.2191 [hep-ph]} \BibitemShut
  {NoStop}%
\bibitem [{\citenamefont {McLerran}\ \emph {et~al.}(2009)\citenamefont
  {McLerran}, \citenamefont {Redlich},\ and\ \citenamefont
  {Sasaki}}]{McLerran:2008ua}%
  \BibitemOpen
  \bibfield  {author} {\bibinfo {author} {\bibfnamefont {L.}~\bibnamefont
  {McLerran}}, \bibinfo {author} {\bibfnamefont {K.}~\bibnamefont {Redlich}}, \
  and\ \bibinfo {author} {\bibfnamefont {C.}~\bibnamefont {Sasaki}},\ }\href
  {\doibase 10.1016/j.nuclphysa.2009.04.001} {\bibfield  {journal} {\bibinfo
  {journal} {Nucl. Phys. A}\ }\textbf {\bibinfo {volume} {824}},\ \bibinfo
  {pages} {86} (\bibinfo {year} {2009})},\ \Eprint
  {http://arxiv.org/abs/0812.3585} {arXiv:0812.3585 [hep-ph]} \BibitemShut
  {NoStop}%
\bibitem [{\citenamefont {Hidaka}\ \emph {et~al.}(2008)\citenamefont {Hidaka},
  \citenamefont {McLerran},\ and\ \citenamefont {Pisarski}}]{Hidaka:2008yy}%
  \BibitemOpen
  \bibfield  {author} {\bibinfo {author} {\bibfnamefont {Y.}~\bibnamefont
  {Hidaka}}, \bibinfo {author} {\bibfnamefont {L.~D.}\ \bibnamefont
  {McLerran}}, \ and\ \bibinfo {author} {\bibfnamefont {R.~D.}\ \bibnamefont
  {Pisarski}},\ }\href {\doibase 10.1016/j.nuclphysa.2008.05.009} {\bibfield
  {journal} {\bibinfo  {journal} {Nucl. Phys. A}\ }\textbf {\bibinfo {volume}
  {808}},\ \bibinfo {pages} {117} (\bibinfo {year} {2008})},\ \Eprint
  {http://arxiv.org/abs/0803.0279} {arXiv:0803.0279 [hep-ph]} \BibitemShut
  {NoStop}%
\bibitem [{\citenamefont {Fukushima}\ and\ \citenamefont
  {Kojo}(2016)}]{Fukushima:2015bda}%
  \BibitemOpen
  \bibfield  {author} {\bibinfo {author} {\bibfnamefont {K.}~\bibnamefont
  {Fukushima}}\ and\ \bibinfo {author} {\bibfnamefont {T.}~\bibnamefont
  {Kojo}},\ }\href {\doibase 10.3847/0004-637X/817/2/180} {\bibfield  {journal}
  {\bibinfo  {journal} {Astrophys. J.}\ }\textbf {\bibinfo {volume} {817}},\
  \bibinfo {pages} {180} (\bibinfo {year} {2016})},\ \Eprint
  {http://arxiv.org/abs/1509.00356} {arXiv:1509.00356 [nucl-th]} \BibitemShut
  {NoStop}%
\bibitem [{\citenamefont {Duarte}\ \emph {et~al.}(2021)\citenamefont {Duarte},
  \citenamefont {Hernandez-Ortiz}, \citenamefont {Jeong},\ and\ \citenamefont
  {McLerran}}]{Duarte:2021tsx}%
  \BibitemOpen
  \bibfield  {author} {\bibinfo {author} {\bibfnamefont {D.~C.}\ \bibnamefont
  {Duarte}}, \bibinfo {author} {\bibfnamefont {S.}~\bibnamefont
  {Hernandez-Ortiz}}, \bibinfo {author} {\bibfnamefont {K.~S.}\ \bibnamefont
  {Jeong}}, \ and\ \bibinfo {author} {\bibfnamefont {L.~D.}\ \bibnamefont
  {McLerran}},\ }\href {\doibase 10.1103/PhysRevD.104.L091901} {\bibfield
  {journal} {\bibinfo  {journal} {Phys. Rev. D}\ }\textbf {\bibinfo {volume}
  {104}},\ \bibinfo {pages} {L091901} (\bibinfo {year} {2021})},\ \Eprint
  {http://arxiv.org/abs/2103.05679} {arXiv:2103.05679 [nucl-th]} \BibitemShut
  {NoStop}%
\bibitem [{\citenamefont {Kojo}(2021)}]{Kojo:2021ugu}%
  \BibitemOpen
  \bibfield  {author} {\bibinfo {author} {\bibfnamefont {T.}~\bibnamefont
  {Kojo}},\ }\href {\doibase 10.1103/PhysRevD.104.074005} {\bibfield  {journal}
  {\bibinfo  {journal} {Phys. Rev. D}\ }\textbf {\bibinfo {volume} {104}},\
  \bibinfo {pages} {074005} (\bibinfo {year} {2021})},\ \Eprint
  {http://arxiv.org/abs/2106.06687} {arXiv:2106.06687 [nucl-th]} \BibitemShut
  {NoStop}%
\bibitem [{\citenamefont {McLerran}\ and\ \citenamefont
  {Reddy}(2019)}]{McLerran:2018hbz}%
  \BibitemOpen
  \bibfield  {author} {\bibinfo {author} {\bibfnamefont {L.}~\bibnamefont
  {McLerran}}\ and\ \bibinfo {author} {\bibfnamefont {S.}~\bibnamefont
  {Reddy}},\ }\href {\doibase 10.1103/PhysRevLett.122.122701} {\bibfield
  {journal} {\bibinfo  {journal} {Phys. Rev. Lett.}\ }\textbf {\bibinfo
  {volume} {122}},\ \bibinfo {pages} {122701} (\bibinfo {year} {2019})},\
  \Eprint {http://arxiv.org/abs/1811.12503} {arXiv:1811.12503 [nucl-th]}
  \BibitemShut {NoStop}%
\bibitem [{\citenamefont {Jeong}\ \emph {et~al.}(2020)\citenamefont {Jeong},
  \citenamefont {McLerran},\ and\ \citenamefont {Sen}}]{Jeong:2019lhv}%
  \BibitemOpen
  \bibfield  {author} {\bibinfo {author} {\bibfnamefont {K.~S.}\ \bibnamefont
  {Jeong}}, \bibinfo {author} {\bibfnamefont {L.}~\bibnamefont {McLerran}}, \
  and\ \bibinfo {author} {\bibfnamefont {S.}~\bibnamefont {Sen}},\ }\href
  {\doibase 10.1103/PhysRevC.101.035201} {\bibfield  {journal} {\bibinfo
  {journal} {Phys. Rev. C}\ }\textbf {\bibinfo {volume} {101}},\ \bibinfo
  {pages} {035201} (\bibinfo {year} {2020})},\ \Eprint
  {http://arxiv.org/abs/1908.04799} {arXiv:1908.04799 [nucl-th]} \BibitemShut
  {NoStop}%
\bibitem [{\citenamefont {Sen}\ and\ \citenamefont
  {Warrington}(2021)}]{Sen:2020peq}%
  \BibitemOpen
  \bibfield  {author} {\bibinfo {author} {\bibfnamefont {S.}~\bibnamefont
  {Sen}}\ and\ \bibinfo {author} {\bibfnamefont {N.~C.}\ \bibnamefont
  {Warrington}},\ }\href {\doibase 10.1016/j.nuclphysa.2020.122059} {\bibfield
  {journal} {\bibinfo  {journal} {Nucl. Phys. A}\ }\textbf {\bibinfo {volume}
  {1006}},\ \bibinfo {pages} {122059} (\bibinfo {year} {2021})},\ \Eprint
  {http://arxiv.org/abs/2002.11133} {arXiv:2002.11133 [nucl-th]} \BibitemShut
  {NoStop}%
\bibitem [{\citenamefont {Fujimoto}\ \emph {et~al.}(2024)\citenamefont
  {Fujimoto}, \citenamefont {Kojo},\ and\ \citenamefont
  {McLerran}}]{Fujimoto:2023mzy}%
  \BibitemOpen
  \bibfield  {author} {\bibinfo {author} {\bibfnamefont {Y.}~\bibnamefont
  {Fujimoto}}, \bibinfo {author} {\bibfnamefont {T.}~\bibnamefont {Kojo}}, \
  and\ \bibinfo {author} {\bibfnamefont {L.~D.}\ \bibnamefont {McLerran}},\
  }\href {\doibase 10.1103/PhysRevLett.132.112701} {\bibfield  {journal}
  {\bibinfo  {journal} {Phys. Rev. Lett.}\ }\textbf {\bibinfo {volume} {132}},\
  \bibinfo {pages} {112701} (\bibinfo {year} {2024})},\ \Eprint
  {http://arxiv.org/abs/2306.04304} {arXiv:2306.04304 [nucl-th]} \BibitemShut
  {NoStop}%
\bibitem [{\citenamefont {Bailin}\ and\ \citenamefont
  {Love}(1984)}]{Bailin:1983bm}%
  \BibitemOpen
  \bibfield  {author} {\bibinfo {author} {\bibfnamefont {D.}~\bibnamefont
  {Bailin}}\ and\ \bibinfo {author} {\bibfnamefont {A.}~\bibnamefont {Love}},\
  }\href {\doibase 10.1016/0370-1573(84)90145-5} {\bibfield  {journal}
  {\bibinfo  {journal} {Phys. Rept.}\ }\textbf {\bibinfo {volume} {107}},\
  \bibinfo {pages} {325} (\bibinfo {year} {1984})}\BibitemShut {NoStop}%
\bibitem [{\citenamefont {Berges}\ and\ \citenamefont
  {Rajagopal}(1999)}]{Berges:1998rc}%
  \BibitemOpen
  \bibfield  {author} {\bibinfo {author} {\bibfnamefont {J.}~\bibnamefont
  {Berges}}\ and\ \bibinfo {author} {\bibfnamefont {K.}~\bibnamefont
  {Rajagopal}},\ }\href {\doibase 10.1016/S0550-3213(98)00620-8} {\bibfield
  {journal} {\bibinfo  {journal} {Nucl. Phys. B}\ }\textbf {\bibinfo {volume}
  {538}},\ \bibinfo {pages} {215} (\bibinfo {year} {1999})},\ \Eprint
  {http://arxiv.org/abs/hep-ph/9804233} {arXiv:hep-ph/9804233} \BibitemShut
  {NoStop}%
\bibitem [{\citenamefont {Rajagopal}\ and\ \citenamefont
  {Wilczek}(2000)}]{Rajagopal:2000wf}%
  \BibitemOpen
  \bibfield  {author} {\bibinfo {author} {\bibfnamefont {K.}~\bibnamefont
  {Rajagopal}}\ and\ \bibinfo {author} {\bibfnamefont {F.}~\bibnamefont
  {Wilczek}},\ }\enquote {\bibinfo {title} {{The Condensed matter physics of
  QCD}},}\ in\ \href {\doibase 10.1142/9789812810458_0043} {\emph {\bibinfo
  {booktitle} {{At the frontier of particle physics. Handbook of QCD. Vol.
  1-3}}}},\ \bibinfo {editor} {edited by\ \bibinfo {editor} {\bibfnamefont
  {M.}~\bibnamefont {Shifman}}\ and\ \bibinfo {editor} {\bibfnamefont
  {B.}~\bibnamefont {Ioffe}}}\ (\bibinfo {year} {2000})\ pp.\ \bibinfo {pages}
  {2061--2151},\ \Eprint {http://arxiv.org/abs/hep-ph/0011333}
  {arXiv:hep-ph/0011333} \BibitemShut {NoStop}%
\bibitem [{\citenamefont {Alford}(2001)}]{Alford:2001dt}%
  \BibitemOpen
  \bibfield  {author} {\bibinfo {author} {\bibfnamefont {M.~G.}\ \bibnamefont
  {Alford}},\ }\href {\doibase 10.1146/annurev.nucl.51.101701.132449}
  {\bibfield  {journal} {\bibinfo  {journal} {Ann. Rev. Nucl. Part. Sci.}\
  }\textbf {\bibinfo {volume} {51}},\ \bibinfo {pages} {131} (\bibinfo {year}
  {2001})},\ \Eprint {http://arxiv.org/abs/hep-ph/0102047}
  {arXiv:hep-ph/0102047} \BibitemShut {NoStop}%
\bibitem [{\citenamefont {Alford}\ \emph {et~al.}(2008)\citenamefont {Alford},
  \citenamefont {Schmitt}, \citenamefont {Rajagopal},\ and\ \citenamefont
  {Sch\"afer}}]{Alford:2007xm}%
  \BibitemOpen
  \bibfield  {author} {\bibinfo {author} {\bibfnamefont {M.~G.}\ \bibnamefont
  {Alford}}, \bibinfo {author} {\bibfnamefont {A.}~\bibnamefont {Schmitt}},
  \bibinfo {author} {\bibfnamefont {K.}~\bibnamefont {Rajagopal}}, \ and\
  \bibinfo {author} {\bibfnamefont {T.}~\bibnamefont {Sch\"afer}},\ }\href
  {\doibase 10.1103/RevModPhys.80.1455} {\bibfield  {journal} {\bibinfo
  {journal} {Rev. Mod. Phys.}\ }\textbf {\bibinfo {volume} {80}},\ \bibinfo
  {pages} {1455} (\bibinfo {year} {2008})},\ \Eprint
  {http://arxiv.org/abs/0709.4635} {arXiv:0709.4635 [hep-ph]} \BibitemShut
  {NoStop}%
\bibitem [{\citenamefont {Detar}\ and\ \citenamefont
  {Kunihiro}(1989)}]{Detar:1988kn}%
  \BibitemOpen
  \bibfield  {author} {\bibinfo {author} {\bibfnamefont {C.~E.}\ \bibnamefont
  {Detar}}\ and\ \bibinfo {author} {\bibfnamefont {T.}~\bibnamefont
  {Kunihiro}},\ }\href {\doibase 10.1103/PhysRevD.39.2805} {\bibfield
  {journal} {\bibinfo  {journal} {Phys. Rev. D}\ }\textbf {\bibinfo {volume}
  {39}},\ \bibinfo {pages} {2805} (\bibinfo {year} {1989})}\BibitemShut
  {NoStop}%
\bibitem [{\citenamefont {Jido}\ \emph {et~al.}(2001)\citenamefont {Jido},
  \citenamefont {Oka},\ and\ \citenamefont {Hosaka}}]{Jido:2001nt}%
  \BibitemOpen
  \bibfield  {author} {\bibinfo {author} {\bibfnamefont {D.}~\bibnamefont
  {Jido}}, \bibinfo {author} {\bibfnamefont {M.}~\bibnamefont {Oka}}, \ and\
  \bibinfo {author} {\bibfnamefont {A.}~\bibnamefont {Hosaka}},\ }\href
  {\doibase 10.1143/PTP.106.873} {\bibfield  {journal} {\bibinfo  {journal}
  {Prog. Theor. Phys.}\ }\textbf {\bibinfo {volume} {106}},\ \bibinfo {pages}
  {873} (\bibinfo {year} {2001})},\ \Eprint
  {http://arxiv.org/abs/hep-ph/0110005} {arXiv:hep-ph/0110005} \BibitemShut
  {NoStop}%
\bibitem [{\citenamefont {Ioffe}(1981)}]{Ioffe:1981kw}%
  \BibitemOpen
  \bibfield  {author} {\bibinfo {author} {\bibfnamefont {B.~L.}\ \bibnamefont
  {Ioffe}},\ }\href {\doibase 10.1016/0550-3213(81)90259-5} {\bibfield
  {journal} {\bibinfo  {journal} {Nucl. Phys. B}\ }\textbf {\bibinfo {volume}
  {188}},\ \bibinfo {pages} {317} (\bibinfo {year} {1981})},\ \bibinfo {note}
  {[Erratum: Nucl.Phys.B 191, 591--592 (1981)]}\BibitemShut {NoStop}%
\bibitem [{\citenamefont {Dominguez}\ and\ \citenamefont
  {de~Rafael}(1987)}]{Dominguez:1986aa}%
  \BibitemOpen
  \bibfield  {author} {\bibinfo {author} {\bibfnamefont {C.~A.}\ \bibnamefont
  {Dominguez}}\ and\ \bibinfo {author} {\bibfnamefont {E.}~\bibnamefont
  {de~Rafael}},\ }\href {\doibase 10.1016/0003-4916(87)90033-9} {\bibfield
  {journal} {\bibinfo  {journal} {Annals Phys.}\ }\textbf {\bibinfo {volume}
  {174}},\ \bibinfo {pages} {372} (\bibinfo {year} {1987})}\BibitemShut
  {NoStop}%
\bibitem [{\citenamefont {Dexheimer}\ \emph {et~al.}(2008)\citenamefont
  {Dexheimer}, \citenamefont {Schramm},\ and\ \citenamefont
  {Zschiesche}}]{Dexheimer:2007tn}%
  \BibitemOpen
  \bibfield  {author} {\bibinfo {author} {\bibfnamefont {V.}~\bibnamefont
  {Dexheimer}}, \bibinfo {author} {\bibfnamefont {S.}~\bibnamefont {Schramm}},
  \ and\ \bibinfo {author} {\bibfnamefont {D.}~\bibnamefont {Zschiesche}},\
  }\href {\doibase 10.1103/PhysRevC.77.025803} {\bibfield  {journal} {\bibinfo
  {journal} {Phys. Rev. C}\ }\textbf {\bibinfo {volume} {77}},\ \bibinfo
  {pages} {025803} (\bibinfo {year} {2008})},\ \Eprint
  {http://arxiv.org/abs/0710.4192} {arXiv:0710.4192 [nucl-th]} \BibitemShut
  {NoStop}%
\bibitem [{\citenamefont {Sasaki}\ and\ \citenamefont
  {Mishustin}(2010)}]{Sasaki:2010bp}%
  \BibitemOpen
  \bibfield  {author} {\bibinfo {author} {\bibfnamefont {C.}~\bibnamefont
  {Sasaki}}\ and\ \bibinfo {author} {\bibfnamefont {I.}~\bibnamefont
  {Mishustin}},\ }\href {\doibase 10.1103/PhysRevC.82.035204} {\bibfield
  {journal} {\bibinfo  {journal} {Phys. Rev. C}\ }\textbf {\bibinfo {volume}
  {82}},\ \bibinfo {pages} {035204} (\bibinfo {year} {2010})},\ \Eprint
  {http://arxiv.org/abs/1005.4811} {arXiv:1005.4811 [hep-ph]} \BibitemShut
  {NoStop}%
\bibitem [{\citenamefont {Motohiro}\ \emph {et~al.}(2015)\citenamefont
  {Motohiro}, \citenamefont {Kim},\ and\ \citenamefont {Harada}}]{Motohiro}%
  \BibitemOpen
  \bibfield  {author} {\bibinfo {author} {\bibfnamefont {Y.}~\bibnamefont
  {Motohiro}}, \bibinfo {author} {\bibfnamefont {Y.}~\bibnamefont {Kim}}, \
  and\ \bibinfo {author} {\bibfnamefont {M.}~\bibnamefont {Harada}},\ }\href
  {\doibase 10.1103/PhysRevC.92.025201} {\bibfield  {journal} {\bibinfo
  {journal} {Phys. Rev. C}\ }\textbf {\bibinfo {volume} {92}},\ \bibinfo
  {pages} {025201} (\bibinfo {year} {2015})}\BibitemShut {NoStop}%
\bibitem [{\citenamefont {Nishihara}\ and\ \citenamefont
  {Harada}(2015)}]{Nishihara:2015fka}%
  \BibitemOpen
  \bibfield  {author} {\bibinfo {author} {\bibfnamefont {H.}~\bibnamefont
  {Nishihara}}\ and\ \bibinfo {author} {\bibfnamefont {M.}~\bibnamefont
  {Harada}},\ }\href {\doibase 10.1103/PhysRevD.92.054022} {\bibfield
  {journal} {\bibinfo  {journal} {Phys. Rev. D}\ }\textbf {\bibinfo {volume}
  {92}},\ \bibinfo {pages} {054022} (\bibinfo {year} {2015})},\ \Eprint
  {http://arxiv.org/abs/1506.07956} {arXiv:1506.07956 [hep-ph]} \BibitemShut
  {NoStop}%
\bibitem [{\citenamefont {Minamikawa}\ \emph
  {et~al.}(2023{\natexlab{b}})\citenamefont {Minamikawa}, \citenamefont {Gao},
  \citenamefont {kojo},\ and\ \citenamefont {Harada}}]{Minamikawa:2023ypn}%
  \BibitemOpen
  \bibfield  {author} {\bibinfo {author} {\bibfnamefont {T.}~\bibnamefont
  {Minamikawa}}, \bibinfo {author} {\bibfnamefont {B.}~\bibnamefont {Gao}},
  \bibinfo {author} {\bibfnamefont {T.}~\bibnamefont {kojo}}, \ and\ \bibinfo
  {author} {\bibfnamefont {M.}~\bibnamefont {Harada}},\ }\href {\doibase
  10.1103/PhysRevD.108.076017} {\bibfield  {journal} {\bibinfo  {journal}
  {Phys. Rev. D}\ }\textbf {\bibinfo {volume} {108}},\ \bibinfo {pages}
  {076017} (\bibinfo {year} {2023}{\natexlab{b}})},\ \Eprint
  {http://arxiv.org/abs/2306.15564} {arXiv:2306.15564 [hep-ph]} \BibitemShut
  {NoStop}%
\bibitem [{\citenamefont {Gao}\ \emph {et~al.}(2024{\natexlab{a}})\citenamefont
  {Gao}, \citenamefont {Kojo},\ and\ \citenamefont {Harada}}]{Gao:2024mew}%
  \BibitemOpen
  \bibfield  {author} {\bibinfo {author} {\bibfnamefont {B.}~\bibnamefont
  {Gao}}, \bibinfo {author} {\bibfnamefont {T.}~\bibnamefont {Kojo}}, \ and\
  \bibinfo {author} {\bibfnamefont {M.}~\bibnamefont {Harada}},\ }\href
  {\doibase 10.1103/PhysRevD.110.016016} {\bibfield  {journal} {\bibinfo
  {journal} {Phys. Rev. D}\ }\textbf {\bibinfo {volume} {110}},\ \bibinfo
  {pages} {016016} (\bibinfo {year} {2024}{\natexlab{a}})},\ \Eprint
  {http://arxiv.org/abs/2403.18214} {arXiv:2403.18214 [hep-ph]} \BibitemShut
  {NoStop}%
\bibitem [{\citenamefont {Hatsuda}\ and\ \citenamefont
  {Prakash}(1989)}]{HATSUDA198911}%
  \BibitemOpen
  \bibfield  {author} {\bibinfo {author} {\bibfnamefont {T.}~\bibnamefont
  {Hatsuda}}\ and\ \bibinfo {author} {\bibfnamefont {M.}~\bibnamefont
  {Prakash}},\ }\href {\doibase https://doi.org/10.1016/0370-2693(89)91040-X}
  {\bibfield  {journal} {\bibinfo  {journal} {Physics Letters B}\ }\textbf
  {\bibinfo {volume} {224}},\ \bibinfo {pages} {11} (\bibinfo {year}
  {1989})}\BibitemShut {NoStop}%
\bibitem [{\citenamefont {Gallas}\ \emph {et~al.}(2011)\citenamefont {Gallas},
  \citenamefont {Giacosa},\ and\ \citenamefont {Pagliara}}]{GALLAS201113}%
  \BibitemOpen
  \bibfield  {author} {\bibinfo {author} {\bibfnamefont {S.}~\bibnamefont
  {Gallas}}, \bibinfo {author} {\bibfnamefont {F.}~\bibnamefont {Giacosa}}, \
  and\ \bibinfo {author} {\bibfnamefont {G.}~\bibnamefont {Pagliara}},\ }\href
  {\doibase https://doi.org/10.1016/j.nuclphysa.2011.09.008} {\bibfield
  {journal} {\bibinfo  {journal} {Nuclear Physics A}\ }\textbf {\bibinfo
  {volume} {872}},\ \bibinfo {pages} {13} (\bibinfo {year} {2011})}\BibitemShut
  {NoStop}%
\bibitem [{\citenamefont {Marczenko}(2024)}]{Marczenko:2024jzn}%
  \BibitemOpen
  \bibfield  {author} {\bibinfo {author} {\bibfnamefont {M.}~\bibnamefont
  {Marczenko}},\ }\href {\doibase 10.1103/PhysRevD.110.014018} {\bibfield
  {journal} {\bibinfo  {journal} {Phys. Rev. D}\ }\textbf {\bibinfo {volume}
  {110}},\ \bibinfo {pages} {014018} (\bibinfo {year} {2024})},\ \Eprint
  {http://arxiv.org/abs/2405.06360} {arXiv:2405.06360 [hep-ph]} \BibitemShut
  {NoStop}%
\bibitem [{\citenamefont {Yasui}\ \emph {et~al.}(2024)\citenamefont {Yasui},
  \citenamefont {Nitta},\ and\ \citenamefont {Sasaki}}]{Yasui:2024dbx}%
  \BibitemOpen
  \bibfield  {author} {\bibinfo {author} {\bibfnamefont {S.}~\bibnamefont
  {Yasui}}, \bibinfo {author} {\bibfnamefont {M.}~\bibnamefont {Nitta}}, \ and\
  \bibinfo {author} {\bibfnamefont {C.}~\bibnamefont {Sasaki}},\ }\href@noop {}
  {\  (\bibinfo {year} {2024})},\ \Eprint {http://arxiv.org/abs/2409.05670}
  {arXiv:2409.05670 [nucl-th]} \BibitemShut {NoStop}%
\bibitem [{\citenamefont {Koch}\ \emph {et~al.}(2024)\citenamefont {Koch},
  \citenamefont {Marczenko}, \citenamefont {Redlich},\ and\ \citenamefont
  {Sasaki}}]{Koch:2023oez}%
  \BibitemOpen
  \bibfield  {author} {\bibinfo {author} {\bibfnamefont {V.}~\bibnamefont
  {Koch}}, \bibinfo {author} {\bibfnamefont {M.}~\bibnamefont {Marczenko}},
  \bibinfo {author} {\bibfnamefont {K.}~\bibnamefont {Redlich}}, \ and\
  \bibinfo {author} {\bibfnamefont {C.}~\bibnamefont {Sasaki}},\ }\href
  {\doibase 10.1103/PhysRevD.109.014033} {\bibfield  {journal} {\bibinfo
  {journal} {Phys. Rev. D}\ }\textbf {\bibinfo {volume} {109}},\ \bibinfo
  {pages} {014033} (\bibinfo {year} {2024})},\ \Eprint
  {http://arxiv.org/abs/2308.15794} {arXiv:2308.15794 [hep-ph]} \BibitemShut
  {NoStop}%
\bibitem [{\citenamefont {Aarts}\ \emph {et~al.}(2015)\citenamefont {Aarts},
  \citenamefont {Allton}, \citenamefont {Hands}, \citenamefont {J\"ager},
  \citenamefont {Praki},\ and\ \citenamefont {Skullerud}}]{Aarts:2015mma}%
  \BibitemOpen
  \bibfield  {author} {\bibinfo {author} {\bibfnamefont {G.}~\bibnamefont
  {Aarts}}, \bibinfo {author} {\bibfnamefont {C.}~\bibnamefont {Allton}},
  \bibinfo {author} {\bibfnamefont {S.}~\bibnamefont {Hands}}, \bibinfo
  {author} {\bibfnamefont {B.}~\bibnamefont {J\"ager}}, \bibinfo {author}
  {\bibfnamefont {C.}~\bibnamefont {Praki}}, \ and\ \bibinfo {author}
  {\bibfnamefont {J.-I.}\ \bibnamefont {Skullerud}},\ }\href {\doibase
  10.1103/PhysRevD.92.014503} {\bibfield  {journal} {\bibinfo  {journal} {Phys.
  Rev. D}\ }\textbf {\bibinfo {volume} {92}},\ \bibinfo {pages} {014503}
  (\bibinfo {year} {2015})},\ \Eprint {http://arxiv.org/abs/1502.03603}
  {arXiv:1502.03603 [hep-lat]} \BibitemShut {NoStop}%
\bibitem [{\citenamefont {Aarts}\ \emph {et~al.}(2017)\citenamefont {Aarts},
  \citenamefont {Allton}, \citenamefont {De~Boni}, \citenamefont {Hands},
  \citenamefont {J\"ager}, \citenamefont {Praki},\ and\ \citenamefont
  {Skullerud}}]{Aarts:2017rrl}%
  \BibitemOpen
  \bibfield  {author} {\bibinfo {author} {\bibfnamefont {G.}~\bibnamefont
  {Aarts}}, \bibinfo {author} {\bibfnamefont {C.}~\bibnamefont {Allton}},
  \bibinfo {author} {\bibfnamefont {D.}~\bibnamefont {De~Boni}}, \bibinfo
  {author} {\bibfnamefont {S.}~\bibnamefont {Hands}}, \bibinfo {author}
  {\bibfnamefont {B.}~\bibnamefont {J\"ager}}, \bibinfo {author} {\bibfnamefont
  {C.}~\bibnamefont {Praki}}, \ and\ \bibinfo {author} {\bibfnamefont {J.-I.}\
  \bibnamefont {Skullerud}},\ }\href {\doibase 10.1007/JHEP06(2017)034}
  {\bibfield  {journal} {\bibinfo  {journal} {JHEP}\ }\textbf {\bibinfo
  {volume} {06}},\ \bibinfo {pages} {034} (\bibinfo {year} {2017})},\ \Eprint
  {http://arxiv.org/abs/1703.09246} {arXiv:1703.09246 [hep-lat]} \BibitemShut
  {NoStop}%
\bibitem [{\citenamefont {Aarts}\ \emph {et~al.}(2018)\citenamefont {Aarts},
  \citenamefont {Allton}, \citenamefont {de~Boni}, \citenamefont {Hands},
  \citenamefont {J\"ager}, \citenamefont {Praki},\ and\ \citenamefont
  {Skullerud}}]{Aarts:2017iai}%
  \BibitemOpen
  \bibfield  {author} {\bibinfo {author} {\bibfnamefont {G.}~\bibnamefont
  {Aarts}}, \bibinfo {author} {\bibfnamefont {C.}~\bibnamefont {Allton}},
  \bibinfo {author} {\bibfnamefont {D.}~\bibnamefont {de~Boni}}, \bibinfo
  {author} {\bibfnamefont {S.}~\bibnamefont {Hands}}, \bibinfo {author}
  {\bibfnamefont {B.}~\bibnamefont {J\"ager}}, \bibinfo {author} {\bibfnamefont
  {C.}~\bibnamefont {Praki}}, \ and\ \bibinfo {author} {\bibfnamefont {J.-I.}\
  \bibnamefont {Skullerud}},\ }\href {\doibase 10.1051/epjconf/201817114005}
  {\bibfield  {journal} {\bibinfo  {journal} {EPJ Web Conf.}\ }\textbf
  {\bibinfo {volume} {171}},\ \bibinfo {pages} {14005} (\bibinfo {year}
  {2018})},\ \Eprint {http://arxiv.org/abs/1710.00566} {arXiv:1710.00566
  [hep-lat]} \BibitemShut {NoStop}%
\bibitem [{\citenamefont {Aarts}\ \emph {et~al.}(2019)\citenamefont {Aarts},
  \citenamefont {Allton}, \citenamefont {De~Boni},\ and\ \citenamefont
  {J\"ager}}]{Aarts:2018glk}%
  \BibitemOpen
  \bibfield  {author} {\bibinfo {author} {\bibfnamefont {G.}~\bibnamefont
  {Aarts}}, \bibinfo {author} {\bibfnamefont {C.}~\bibnamefont {Allton}},
  \bibinfo {author} {\bibfnamefont {D.}~\bibnamefont {De~Boni}}, \ and\
  \bibinfo {author} {\bibfnamefont {B.}~\bibnamefont {J\"ager}},\ }\href
  {\doibase 10.1103/PhysRevD.99.074503} {\bibfield  {journal} {\bibinfo
  {journal} {Phys. Rev. D}\ }\textbf {\bibinfo {volume} {99}},\ \bibinfo
  {pages} {074503} (\bibinfo {year} {2019})},\ \Eprint
  {http://arxiv.org/abs/1812.07393} {arXiv:1812.07393 [hep-lat]} \BibitemShut
  {NoStop}%
\bibitem [{\citenamefont {Marczenko}\ \emph {et~al.}(2019)\citenamefont
  {Marczenko}, \citenamefont {Blaschke}, \citenamefont {Redlich},\ and\
  \citenamefont {Sasaki}}]{universe5080180}%
  \BibitemOpen
  \bibfield  {author} {\bibinfo {author} {\bibfnamefont {M.}~\bibnamefont
  {Marczenko}}, \bibinfo {author} {\bibfnamefont {D.}~\bibnamefont {Blaschke}},
  \bibinfo {author} {\bibfnamefont {K.}~\bibnamefont {Redlich}}, \ and\
  \bibinfo {author} {\bibfnamefont {C.}~\bibnamefont {Sasaki}},\ }\href
  {\doibase 10.3390/universe5080180} {\bibfield  {journal} {\bibinfo  {journal}
  {Universe}\ }\textbf {\bibinfo {volume} {5}} (\bibinfo {year} {2019}),\
  10.3390/universe5080180}\BibitemShut {NoStop}%
\bibitem [{\citenamefont {Yamazaki}\ and\ \citenamefont
  {Harada}(2019)}]{PhysRevC.100.025205}%
  \BibitemOpen
  \bibfield  {author} {\bibinfo {author} {\bibfnamefont {T.}~\bibnamefont
  {Yamazaki}}\ and\ \bibinfo {author} {\bibfnamefont {M.}~\bibnamefont
  {Harada}},\ }\href {\doibase 10.1103/PhysRevC.100.025205} {\bibfield
  {journal} {\bibinfo  {journal} {Phys. Rev. C}\ }\textbf {\bibinfo {volume}
  {100}},\ \bibinfo {pages} {025205} (\bibinfo {year} {2019})}\BibitemShut
  {NoStop}%
\bibitem [{\citenamefont {Mukherjee}\ \emph {et~al.}(2017)\citenamefont
  {Mukherjee}, \citenamefont {Schramm}, \citenamefont {Steinheimer},\ and\
  \citenamefont {Dexheimer}}]{Mukherjee:2017jzi}%
  \BibitemOpen
  \bibfield  {author} {\bibinfo {author} {\bibfnamefont {A.}~\bibnamefont
  {Mukherjee}}, \bibinfo {author} {\bibfnamefont {S.}~\bibnamefont {Schramm}},
  \bibinfo {author} {\bibfnamefont {J.}~\bibnamefont {Steinheimer}}, \ and\
  \bibinfo {author} {\bibfnamefont {V.}~\bibnamefont {Dexheimer}},\ }\href
  {\doibase 10.1051/0004-6361/201731505} {\bibfield  {journal} {\bibinfo
  {journal} {Astron. Astrophys.}\ }\textbf {\bibinfo {volume} {608}},\ \bibinfo
  {pages} {A110} (\bibinfo {year} {2017})},\ \Eprint
  {http://arxiv.org/abs/1706.09191} {arXiv:1706.09191 [nucl-th]} \BibitemShut
  {NoStop}%
\bibitem [{\citenamefont {Minamikawa}\ \emph {et~al.}(2021)\citenamefont
  {Minamikawa}, \citenamefont {Kojo},\ and\ \citenamefont
  {Harada}}]{Minamikawa:2020jfj}%
  \BibitemOpen
  \bibfield  {author} {\bibinfo {author} {\bibfnamefont {T.}~\bibnamefont
  {Minamikawa}}, \bibinfo {author} {\bibfnamefont {T.}~\bibnamefont {Kojo}}, \
  and\ \bibinfo {author} {\bibfnamefont {M.}~\bibnamefont {Harada}},\ }\href
  {\doibase 10.1103/PhysRevC.103.045205} {\bibfield  {journal} {\bibinfo
  {journal} {Phys. Rev. C}\ }\textbf {\bibinfo {volume} {103}},\ \bibinfo
  {pages} {045205} (\bibinfo {year} {2021})},\ \Eprint
  {http://arxiv.org/abs/2011.13684} {arXiv:2011.13684 [nucl-th]} \BibitemShut
  {NoStop}%
\bibitem [{\citenamefont {Marczenko}\ \emph
  {et~al.}(2022{\natexlab{a}})\citenamefont {Marczenko}, \citenamefont
  {Redlich},\ and\ \citenamefont {Sasaki}}]{Marczenko:2021uaj}%
  \BibitemOpen
  \bibfield  {author} {\bibinfo {author} {\bibfnamefont {M.}~\bibnamefont
  {Marczenko}}, \bibinfo {author} {\bibfnamefont {K.}~\bibnamefont {Redlich}},
  \ and\ \bibinfo {author} {\bibfnamefont {C.}~\bibnamefont {Sasaki}},\ }\href
  {\doibase 10.3847/2041-8213/ac4b61} {\bibfield  {journal} {\bibinfo
  {journal} {Astrophys. J. Lett.}\ }\textbf {\bibinfo {volume} {925}},\
  \bibinfo {pages} {L23} (\bibinfo {year} {2022}{\natexlab{a}})},\ \Eprint
  {http://arxiv.org/abs/2110.11056} {arXiv:2110.11056 [nucl-th]} \BibitemShut
  {NoStop}%
\bibitem [{\citenamefont {Marczenko}\ \emph
  {et~al.}(2022{\natexlab{b}})\citenamefont {Marczenko}, \citenamefont
  {Redlich},\ and\ \citenamefont {Sasaki}}]{Marczenko:2022hyt}%
  \BibitemOpen
  \bibfield  {author} {\bibinfo {author} {\bibfnamefont {M.}~\bibnamefont
  {Marczenko}}, \bibinfo {author} {\bibfnamefont {K.}~\bibnamefont {Redlich}},
  \ and\ \bibinfo {author} {\bibfnamefont {C.}~\bibnamefont {Sasaki}},\ }\href
  {\doibase 10.1103/PhysRevD.105.103009} {\bibfield  {journal} {\bibinfo
  {journal} {Phys. Rev. D}\ }\textbf {\bibinfo {volume} {105}},\ \bibinfo
  {pages} {103009} (\bibinfo {year} {2022}{\natexlab{b}})},\ \Eprint
  {http://arxiv.org/abs/2203.00269} {arXiv:2203.00269 [nucl-th]} \BibitemShut
  {NoStop}%
\bibitem [{\citenamefont {Kong}\ \emph {et~al.}(2023)\citenamefont {Kong},
  \citenamefont {Minamikawa},\ and\ \citenamefont {Harada}}]{Kong:2023nue}%
  \BibitemOpen
  \bibfield  {author} {\bibinfo {author} {\bibfnamefont {Y.~K.}\ \bibnamefont
  {Kong}}, \bibinfo {author} {\bibfnamefont {T.}~\bibnamefont {Minamikawa}}, \
  and\ \bibinfo {author} {\bibfnamefont {M.}~\bibnamefont {Harada}},\ }\href
  {\doibase 10.1103/PhysRevC.108.055206} {\bibfield  {journal} {\bibinfo
  {journal} {Phys. Rev. C}\ }\textbf {\bibinfo {volume} {108}},\ \bibinfo
  {pages} {055206} (\bibinfo {year} {2023})},\ \Eprint
  {http://arxiv.org/abs/2306.08140} {arXiv:2306.08140 [nucl-th]} \BibitemShut
  {NoStop}%
\bibitem [{\citenamefont {Gao}\ \emph {et~al.}(2024{\natexlab{b}})\citenamefont
  {Gao}, \citenamefont {Yan},\ and\ \citenamefont {Harada}}]{Gao:2024chh}%
  \BibitemOpen
  \bibfield  {author} {\bibinfo {author} {\bibfnamefont {B.}~\bibnamefont
  {Gao}}, \bibinfo {author} {\bibfnamefont {Y.}~\bibnamefont {Yan}}, \ and\
  \bibinfo {author} {\bibfnamefont {M.}~\bibnamefont {Harada}},\ }\href
  {\doibase 10.1103/PhysRevC.109.065807} {\bibfield  {journal} {\bibinfo
  {journal} {Phys. Rev. C}\ }\textbf {\bibinfo {volume} {109}},\ \bibinfo
  {pages} {065807} (\bibinfo {year} {2024}{\natexlab{b}})},\ \Eprint
  {http://arxiv.org/abs/2404.04786} {arXiv:2404.04786 [nucl-th]} \BibitemShut
  {NoStop}%
\bibitem [{\citenamefont {Gao}\ \emph {et~al.}(2024{\natexlab{c}})\citenamefont
  {Gao}, \citenamefont {Yuan}, \citenamefont {Harada},\ and\ \citenamefont
  {Ma}}]{Gao:2024lzu}%
  \BibitemOpen
  \bibfield  {author} {\bibinfo {author} {\bibfnamefont {B.}~\bibnamefont
  {Gao}}, \bibinfo {author} {\bibfnamefont {W.-L.}\ \bibnamefont {Yuan}},
  \bibinfo {author} {\bibfnamefont {M.}~\bibnamefont {Harada}}, \ and\ \bibinfo
  {author} {\bibfnamefont {Y.-L.}\ \bibnamefont {Ma}},\ }\href {\doibase
  10.1103/PhysRevC.110.045802} {\bibfield  {journal} {\bibinfo  {journal}
  {Phys. Rev. C}\ }\textbf {\bibinfo {volume} {110}},\ \bibinfo {pages}
  {045802} (\bibinfo {year} {2024}{\natexlab{c}})},\ \Eprint
  {http://arxiv.org/abs/2407.13990} {arXiv:2407.13990 [nucl-th]} \BibitemShut
  {NoStop}%
\bibitem [{\citenamefont {Eser}\ and\ \citenamefont
  {Blaizot}(2024)}]{Eser:2024xil}%
  \BibitemOpen
  \bibfield  {author} {\bibinfo {author} {\bibfnamefont {J.}~\bibnamefont
  {Eser}}\ and\ \bibinfo {author} {\bibfnamefont {J.-P.}\ \bibnamefont
  {Blaizot}},\ }\href@noop {} {\  (\bibinfo {year} {2024})},\ \Eprint
  {http://arxiv.org/abs/2408.01302} {arXiv:2408.01302 [nucl-th]} \BibitemShut
  {NoStop}%
\bibitem [{\citenamefont {De~Rujula}\ \emph {et~al.}(1975)\citenamefont
  {De~Rujula}, \citenamefont {Georgi},\ and\ \citenamefont
  {Glashow}}]{DeRujula:1975qlm}%
  \BibitemOpen
  \bibfield  {author} {\bibinfo {author} {\bibfnamefont {A.}~\bibnamefont
  {De~Rujula}}, \bibinfo {author} {\bibfnamefont {H.}~\bibnamefont {Georgi}}, \
  and\ \bibinfo {author} {\bibfnamefont {S.~L.}\ \bibnamefont {Glashow}},\
  }\href {\doibase 10.1103/PhysRevD.12.147} {\bibfield  {journal} {\bibinfo
  {journal} {Phys. Rev. D}\ }\textbf {\bibinfo {volume} {12}},\ \bibinfo
  {pages} {147} (\bibinfo {year} {1975})}\BibitemShut {NoStop}%
\bibitem [{\citenamefont {Isgur}\ and\ \citenamefont
  {Karl}(1978)}]{Isgur:1978xj}%
  \BibitemOpen
  \bibfield  {author} {\bibinfo {author} {\bibfnamefont {N.}~\bibnamefont
  {Isgur}}\ and\ \bibinfo {author} {\bibfnamefont {G.}~\bibnamefont {Karl}},\
  }\href {\doibase 10.1103/PhysRevD.18.4187} {\bibfield  {journal} {\bibinfo
  {journal} {Phys. Rev. D}\ }\textbf {\bibinfo {volume} {18}},\ \bibinfo
  {pages} {4187} (\bibinfo {year} {1978})}\BibitemShut {NoStop}%
\bibitem [{\citenamefont {Manohar}\ and\ \citenamefont
  {Georgi}(1984)}]{Manohar:1983md}%
  \BibitemOpen
  \bibfield  {author} {\bibinfo {author} {\bibfnamefont {A.}~\bibnamefont
  {Manohar}}\ and\ \bibinfo {author} {\bibfnamefont {H.}~\bibnamefont
  {Georgi}},\ }\href {\doibase 10.1016/0550-3213(84)90231-1} {\bibfield
  {journal} {\bibinfo  {journal} {Nucl. Phys. B}\ }\textbf {\bibinfo {volume}
  {234}},\ \bibinfo {pages} {189} (\bibinfo {year} {1984})}\BibitemShut
  {NoStop}%
\bibitem [{\citenamefont {Scavenius}\ \emph {et~al.}(2001)\citenamefont
  {Scavenius}, \citenamefont {Mocsy}, \citenamefont {Mishustin},\ and\
  \citenamefont {Rischke}}]{Scavenius:2000qd}%
  \BibitemOpen
  \bibfield  {author} {\bibinfo {author} {\bibfnamefont {O.}~\bibnamefont
  {Scavenius}}, \bibinfo {author} {\bibfnamefont {A.}~\bibnamefont {Mocsy}},
  \bibinfo {author} {\bibfnamefont {I.~N.}\ \bibnamefont {Mishustin}}, \ and\
  \bibinfo {author} {\bibfnamefont {D.~H.}\ \bibnamefont {Rischke}},\ }\href
  {\doibase 10.1103/PhysRevC.64.045202} {\bibfield  {journal} {\bibinfo
  {journal} {Phys. Rev. C}\ }\textbf {\bibinfo {volume} {64}},\ \bibinfo
  {pages} {045202} (\bibinfo {year} {2001})},\ \Eprint
  {http://arxiv.org/abs/nucl-th/0007030} {arXiv:nucl-th/0007030} \BibitemShut
  {NoStop}%
\bibitem [{\citenamefont {Fischer}\ \emph {et~al.}(2004)\citenamefont
  {Fischer}, \citenamefont {Alkofer}, \citenamefont {Dahm},\ and\ \citenamefont
  {Maris}}]{Fischer:2004nq}%
  \BibitemOpen
  \bibfield  {author} {\bibinfo {author} {\bibfnamefont {C.~S.}\ \bibnamefont
  {Fischer}}, \bibinfo {author} {\bibfnamefont {R.}~\bibnamefont {Alkofer}},
  \bibinfo {author} {\bibfnamefont {T.}~\bibnamefont {Dahm}}, \ and\ \bibinfo
  {author} {\bibfnamefont {P.}~\bibnamefont {Maris}},\ }\href {\doibase
  10.1103/PhysRevD.70.073007} {\bibfield  {journal} {\bibinfo  {journal} {Phys.
  Rev. D}\ }\textbf {\bibinfo {volume} {70}},\ \bibinfo {pages} {073007}
  (\bibinfo {year} {2004})},\ \Eprint {http://arxiv.org/abs/hep-ph/0407104}
  {arXiv:hep-ph/0407104} \BibitemShut {NoStop}%
\bibitem [{\citenamefont {Greensite}(2017)}]{Greensite:2016pfc}%
  \BibitemOpen
  \bibfield  {author} {\bibinfo {author} {\bibfnamefont {J.}~\bibnamefont
  {Greensite}},\ }\href {\doibase 10.1051/epjconf/201713701009} {\bibfield
  {journal} {\bibinfo  {journal} {EPJ Web Conf.}\ }\textbf {\bibinfo {volume}
  {137}},\ \bibinfo {pages} {01009} (\bibinfo {year} {2017})},\ \Eprint
  {http://arxiv.org/abs/1610.06221} {arXiv:1610.06221 [hep-lat]} \BibitemShut
  {NoStop}%
\bibitem [{\citenamefont {Mazur}\ \emph {et~al.}(2023)\citenamefont {Mazur},
  \citenamefont {Kim}, \citenamefont {Harada},\ and\ \citenamefont
  {Lee}}]{Mazur:2020anw}%
  \BibitemOpen
  \bibfield  {author} {\bibinfo {author} {\bibfnamefont {I.~A.}\ \bibnamefont
  {Mazur}}, \bibinfo {author} {\bibfnamefont {Y.}~\bibnamefont {Kim}}, \bibinfo
  {author} {\bibfnamefont {M.}~\bibnamefont {Harada}}, \ and\ \bibinfo {author}
  {\bibfnamefont {H.~K.}\ \bibnamefont {Lee}},\ }\href {\doibase
  10.1142/S021830132350060X} {\bibfield  {journal} {\bibinfo  {journal} {Int.
  J. Mod. Phys. E}\ }\textbf {\bibinfo {volume} {32}},\ \bibinfo {pages}
  {2350060} (\bibinfo {year} {2023})},\ \Eprint
  {http://arxiv.org/abs/2009.02839} {arXiv:2009.02839 [hep-ph]} \BibitemShut
  {NoStop}%
\end{thebibliography}%

\end{document}